\documentclass[12pt]{article}
\usepackage{epsfig,amssymb}

\newcommand{\bq}{\begin{equation}}
\newcommand{\eq}{\end{equation}}
\newcommand{\bqa}{\begin{eqnarray}}
\newcommand{\eqa}{\end{eqnarray}}
\newcommand{\ben}{\begin{enumerate}}
\newcommand{\een}{\end{enumerate}}
\newcommand{\bc}{\begin{center}}
\newcommand{\ec}{\end{center}}
\newcommand{\bqb}{\begin{eqnarray*}}
\newcommand{\eqb}{\end{eqnarray*}}

\def\pr#1#2#3{ Phys. Rev. ${\bf{#1}}$, #2 (#3)}
\def\prl#1#2#3{ Phys. Rev. Lett. ${\bf{#1}}$, #2 (#3)}
\def\pl#1#2#3{ Phys. Lett. ${\bf{#1}}$, #2 (#3)}
\def\prep#1#2#3{ Phys. Rep. ${\bf{#1}}$, #2 (#3)}

\def\np#1#2#3{ Nucl. Phys. ${\bf{#1}}$, #2 (#3)}
\def\zp#1#2#3{ Z. f. Phys. ${\bf{#1}}$, #2 (#3)}
\def\epj#1#2#3{ Eur. Phys. J. ${\bf{#1}}$, #2 (#3)}

\def\cpc#1#2#3{Comput. Phys. Commun. ${\bf{#1}}$, #2 (#3)}

\def\ie{{\it i.e.\/}}
\def\eg{{\it e.g.\/}}

\def\etal{{\it et.al.\/}}

\global\nulldelimiterspace = 0pt

\def\L{ {\cal L }}

\def\A{ {\cal A }}

\def\V{ {\cal V }}
\def\sw{s_W}
\def\cw{c_W}
\def\swd{s^2_W}
\def\cwd{c^2_W}

\def\mwd{m_W^2}
\def\mw{m_W}
\def\mz{m_Z}
\def\mzd{m_Z^2}

\def\LamNP{\Lambda_{NP}}

\def\Sn#1{\mathrm{Sign} #1 }

\begin{document}
\pagenumbering{arabic}
\thispagestyle{empty}
\def\thefootnote{\fnsymbol{footnote}}
\setcounter{footnote}{1}

\begin{flushright}
PM/00-11 \\
THES-TP 2000/01 \\
March 2000\\

 \end{flushright}
\vspace{2cm}
\begin{center}
{\Large\bf New and Standard Physics contributions to anomalous $Z$ and
$\gamma$ self-couplings}\footnote{Partially supported by the European
Community grant ERBFMRX-CT96-0090.}
 \vspace{1.5cm}  \\
{\large G.J. Gounaris$^a$, J. Layssac$^b$  and F.M. Renard$^b$}\\
\vspace{0.7cm}
$^a$Department of Theoretical Physics, Aristotle
University of Thessaloniki,\\
Gr-54006, Thessaloniki, Greece.\\
\vspace{0.2cm}
$^b$Physique
Math\'{e}matique et Th\'{e}orique,
UMR 5825\\
Universit\'{e} Montpellier II,
 F-34095 Montpellier Cedex 5.\\
\vspace{0.2cm}

\vspace*{1cm}

{\bf Abstract}
\end{center}

We examine the Standard  and the  New Physics (NP) contributions
to the   $ZZZ$, $ZZ\gamma$ and $Z\gamma\gamma$
neutral gauge couplings. At the one-loop level, if we assume
that there is no CP violation contained in NP
beyond the  Standard Model one,   we find that
only CP conserving  neutral gauge couplings
are  generated,  either from the standard quarks and leptons, or
from possible New Physics (NP) fermions.
Bosonic one-loop diagrams never contribute to these
couplings, while the aforementioned  fermionic contributions satisfy
$h^Z_3 \simeq - f^{\gamma}_5$, $h^Z_4=h^{\gamma}_4=0$.
We also study   examples of two-loop NP effects that
could generate non vanishing $h^{\gamma,Z}_4$ couplings.
We compare quantitative estimates from SM, MSSM and
some specific examples of NP contributions, and we discuss their
observability at future colliders.\par

\def\thefootnote{\arabic{footnote}}
\setcounter{footnote}{0}
\clearpage

\section{Introduction}

Recently there has been a renewed interest in the possible existence of
anomalous neutral gauge boson self-couplings. This is due to the
acquisition  of new experimental results at LEP2 \cite{LEP2ZZ} which,
together with the TEVATRON  results  \cite{TEVZg}, begin to
produce interesting constraints on such couplings; which should
further   improve in the future at the next colliders \cite{Alcaraz}.

This has lead to a reexamination of the phenomenological
description commonly used for these couplings. The necessity of
certain corrections was discovered  and their
implications for $ZZ$ and $Z\gamma$ production at $e^-e^+$
and hadron colliders were  discussed  \cite{neut}.\par

For what concerns the quantitative theoretical predictions for each
of these neutral couplings, very little has been said up to now
\cite{FMR, BBCD, Baur, Boudjema}, in  contrary to the
of the charged ($ZWW$ and $\gamma WW$) self-couplings
for which several types of
predictions had been given since a long time \cite{Wudka, work, Schi}.
A reappraisal of the theoretical expectations for
these neutral couplings, is still lacking.
Therefore, the purpose of this paper is to fill this lack
and study the Standard Model (SM) predictions for these couplings,
as well as the  predictions arising
from possible new physics beyond it.   \par

In Section 2 we first recall some general properties following from
Bose statistics, Lorentz symmetry  and $SU(2)\times
U(1)$ gauge invariant effective lagrangians. The most notable of
them  is that the neutral gauge couplings  vanish
whenever all three gauge boson are on-shell. Thus,
at least one the gauge boson need to be off-shell,
for such couplings to appear.  Then, in Section 3
we consider the perturbative contributions to these couplings
arising  at one loop. When standard  vertices
for the gauge boson interactions are used, and in particular
no CP violation in  the photon- and $Z$- couplings to
 fermions is considered, then of course only CP conserving  neutral
gauge self-couplings can arise. At the 1-loop level,
such couplings can only be induced by a  fermionic triangle diagram,
involving either new or SM  fermions.
We give the exact expression of these
contributions to the real and  imaginary parts of the neutral gauge
 couplings,
in terms of the fermionic ones  $(g_{vj},~ g_{aj})$
and  the fermion masses $M_j$, as well as  the
 squared mass $s$ of the off-shell vector boson.
To elucidate the remarkable properties of these results, we study both
the high $s$  behaviour  of these gauge couplings at
fixed  fermion  mass $M_j$,  as well as their
high fermion-mass limit ($M_j^2 \gg s,~ m^2_Z$).
As we will see, the behaviour in these limits is intimately
related to the  way the anomaly cancellation is realized. \par

The quantitative aspects of these  fermionic contributions are
discussed in Section 4, where we consider   the  SM
contributions to the real and  imaginary parts of the
couplings, as well as the relative magnitude of the lepton- ,
 light quark- and  the top quark-contributions. We observe that
the anomaly cancellation is intimately accompanied by
considerable cancellation between the lepton and quark
contribution to the physical neutral gauge boson couplings at high
energy. We then consider the
supersymmetric contributions due to the charginos and  neutralinos
in the MSSM. And finally we discuss the possibility of heavy  fermions
associated to some form of new physics (NP) with a high intrinsic
scale $\Lambda_{NP}$.\par

Section 5 is devoted to contributions that could arise beyond the
fermionic one loop level, either though higher order
perturbative diagrams or through  non perturbative effects.
Finally, in   Section 6, we summarize our results and their consequences
for the observability of neutral self-boson couplings at present and
future colliders.

\section{General properties of neutral self-boson couplings}

\begin{figure}[p]
\vspace*{-4cm}
\[
\epsfig{file=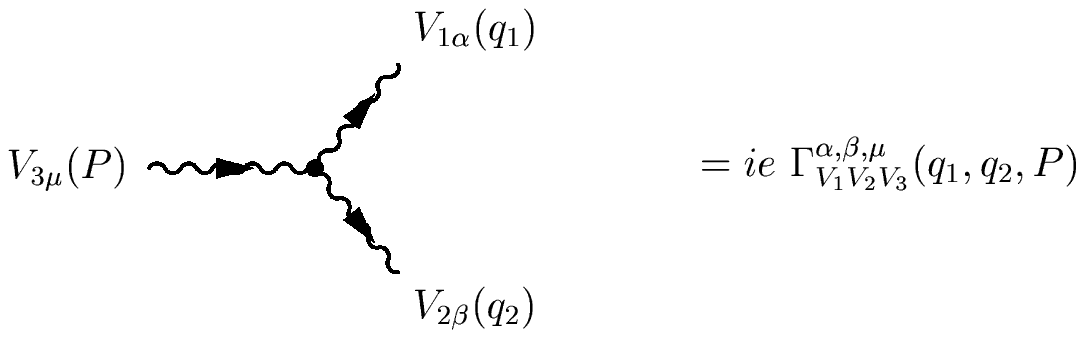,height=3cm,width=12cm}
\]
\vspace*{0.5cm}
\caption[1]{The general neutral gauge boson vertex
 $V_1V_2V_3$.}
\label{Feyn-fig}
\end{figure}

 Because of Bose statistics, the $Z$ and $\gamma$ self-couplings
 vanish identically, when all three particles are on-shell.
The general form of the couplings of one off-shell boson
($V=Z,\gamma$) to a final pair of
on-shell $ZZ$ or  $Z\gamma$ bosons, is
\bqa
\Gamma^{\alpha \beta \mu}_{ZZ V} (q_1, q_2, P)
&=& \frac{i (s-m_V^2)}{\mzd}
\left [ f_4^V (P^\alpha g^{\mu \beta}+P^\beta g^{\mu \alpha})
-f_5^V \epsilon^{\mu \alpha \beta \rho}(q_1-q_2)_\rho \right ]
~, \label{fZZ} \\
\Gamma^{\alpha \beta \mu}_{Z\gamma V} (q_1, q_2, P)
&=& \frac{ i (s-m_V^2)}{\mzd}
\Bigg \{ h_1^V (q_2^\mu g^{\alpha \beta}-q_2^\alpha g^{\mu \beta} )
+ \frac{h_2^V}{\mzd} P^\alpha [ (Pq_2) g^{\mu \beta}- q_2^\mu P^\beta ]
\nonumber \\
&-& h_3^V \epsilon^{\mu \alpha \beta \rho} q_{2\rho}
~-~\frac{h_4^V}{\mzd} P^\alpha \epsilon^{\mu \beta \rho
\sigma}P_\rho q_{2\sigma} \Bigg \}~ , \label{hZgamma}
\eqa
where the momenta are defined as in Fig.\ref{Feyn-fig} and $s\equiv P^2$,
is\footnote{$\epsilon^{0123}=1$.} used.
The expressions (\ref{fZZ}, \ref{hZgamma})
follow from the general forms written in
 \cite{Hagiwara, GG-TGC} and the corrections made
in \cite{neut}. The forms associated to $f_4^V,~h_1^V,~h_2^V$ are
CP-violating, whereas the ones associated to $f_5^V,~h_3^V,~h_4^V$
are CP-conserving.\par

The CP-conserving forms in (\ref{fZZ}, \ref{hZgamma})
are C- and P- violating and in this respect
 they are analogous to the anapole $ZW^+W^-$ and $\gamma W^+W^-$ vertices
\bq
\Gamma^{\alpha \beta \mu}_{W^+W^- V} (q_1, q_2, P)
=i{z_V\over m^2_W}\Bigg \{\epsilon^{\mu \alpha \sigma \rho}
P_{\sigma}(q_1-q_2)_\rho P^{\beta}-\epsilon^{\mu \beta \sigma \rho}
P_{\sigma}(q_1-q_2)_\rho P^{\alpha}\Bigg \}~ , \label{ZWW-anapole}
\eq
as well to the corresponding gauge boson-fermion anomalous anapole
coupling.
None of these couplings exist at tree level in the Standard Model (SM).
At the one-loop SM level though, as we will see below, such couplings
do appear and tend to be strongly decreasing with  $s$.\par

Since the CP violating couplings in (\ref{fZZ}, \ref{hZgamma})
can never be generated, if the NP interactions
of $Z$ and photon conserve CP,  we concentrate below on the CP
conserving couplings $f_5^V,~h_3^V,~h_4^V$. As already observed
these are analogous to the anapole ones. But the situation in this
 neutral anapole sector is rather different
from the one in the  sector of the general charged $ZWW$ and
$\gamma WW$ couplings. This can   been seen by comparing
the  results of the calculation of the triangular graph of
Fig.\ref{trif-fig}, with the generic expectations from
a dimensional analysis in the effective lagrangian framework.
More explicitly, the contribution of a heavy fermion of mass
$\LamNP$  to the aforementioned 1-loop triangular graph results to
an  $f_5^V$ or $h_3^V$ coupling, which may occasionally behave like
$(\mw/\LamNP)^2$. On the other hand, when one writes the effective
lagrangian in terms of $SU(2)\times U(1)$
gauge invariant operators in the linear representation
\cite{Buchmuller}, then at the lowest non-trivial level of
$dim=6$ operators several anomalous $ZWW$ and $\gamma WW$
couplings are generated  \cite{Hagiwara2, RG-effective}. However,
at this level, neither the anapole $ZWW$ coupling
of  (\ref{ZWW-anapole}), nor any  neutral gauge couplings ever
appear. These couplings require higher dimensional
$dim \geq 8$ operators, which means that their magnitude
should be depressed by at least one more power of
$\mw^2/\LamNP^2$ and behave like\footnote{The same conclusion
should also be valid  if the non-linear
Higgs representation is used. In this later case
the $ZWW$ anapole coupling can be generated
at the dominant $D_{chiral}=4$ level; but the generation of
 neutral self-couplings still requires higher
dimensional operators \cite{Yuan}.}
 $(m_W/\Lambda_{NP})^4$. \par

It is therefore interesting to examine more precisely the
conditions under which such couplings can  be
generated and what type of NP effects determine their magnitude.
\par

\section{Fermion loop contributions}

We have first looked at the perturbative ways in which the neutral
couplings in (\ref{fZZ}, \ref{hZgamma}) could be generated.
One immediately observes that at the 1-loop level
the relevant graphs are triangular ones  of  the type of
Fig.\ref{trif-fig}. For  scalars or  $W^{\pm}$ bosons running
along the loop in such graphs, with standard $ZWW$
and $\gamma WW$ couplings,  we always get identically vanishing
contributions.
In particular for the CP conserving couplings, the reason is
that the   $\epsilon^{\mu\nu\rho\sigma}$ tensor can never be
generated from them. Only a fermionic loop (either with a single
fermion $F_j$ running along the loop, or with mixed
$F_1,~ F_2,~... $ fermionic  contributions), can generate
such  $\epsilon^{\mu\nu\rho\sigma}$ terms,
through the axial $Z$ coupling (see Fig.\ref{trif-fig}).
To describe them, we use the standard definitions
\bqa
\L & = & -eQ_j A^\mu\bar F_j \gamma_\mu F_j
-{e\over2s_Wc_W}  Z^\mu \bar F_j \left  ( \gamma_\mu g_{vj}
- \gamma_\mu \gamma_5 g_{aj} \right ) F_j
\nonumber \\
&& -~{e\over2s_Wc_W}  Z^\mu \bar F_1 \left  ( \gamma_\mu g_{v12}
- \gamma_\mu \gamma_5 g_{a12} \right ) F_2
  ~~ , \label{VheavyF}
\eqa
\noindent
where $Q_j$ is the $F_j$ charge, while $g_{vj}$, $g_{aj}$
and the mixed couplings  $g_{v12}$, $g_{a12}$ determine the $Z$-fermion
interactions. If there are no CP violating NP sources, then all these
couplings must be real, and hermiticity requires
$g_{v12}=g_{v21}$, $g_{a12}=g_{a21}$. As already said, in such a case
only the CP conserving neutral gauge boson couplings
 in (\ref{fZZ}, \ref{hZgamma}) can in principle be generated.

Using standard techniques for  preserving CVC and
Bose symmetry, (or equivalently for isolating the anomaly
contribution) we  get the 1-loop
fermionic contributions  to the CP conserving couplings
 $h^{Z,\gamma}_3$ and $f^{Z,\gamma}_5$ presented
in Appendix A in terms of
Passarino-Veltman functions, and in Appendix B in terms of the Feynman
parametrization. These expressions are  used below for computing
the precise predictions of the Standard Model and of the MSSM.
Before presenting  these though, the following remarks
are in order. \par

We first note that at the one-loop level, we can never generate
 $h_4^Z$ and $h_4^\gamma$, \ie~
\bq  h^Z_4\equiv h^{\gamma}_4\equiv 0 ~ . \label{h4Zg}
\eq
Therefore, the
neutral couplings most likely to appear are $f_5^{\gamma, Z}$ and
$h_3^{\gamma ,Z}$. From the expressions given in the
Appendices A or B, it is easy  to obtain their behaviour
in the high and low energy limit.
At  \underline{high energy} ~ $s \gg  M^2_j,~m^2_Z$,  we get
\bqa
 h^Z_3& \simeq  &-f^{\gamma}_5\simeq
~N_F{e^2Q_jg_{vj}g_{aj}\over8\pi^2s^2_Wc^2_W}~
\left ({m^2_Z\over s}\right ) ~ ,\label{h3Zs} \\
f^{Z}_5& \simeq & -~N_F{e^2g_{aj}[g_{aj}^2+3g^2_{vj}]
\over48\pi^2s^3_Wc^3_W}~ \left ({m^2_Z\over s}\right ) ~ ,
\label{f5Zs} \\
h^{\gamma}_3& \simeq &~N_F{e^2Q_jg_{vj}g_{aj}\over4\pi^2s_Wc_W}~
\left ({m^2_Z\over s}\right ) ~ ,\label{h3gs}
\eqa
where $M_j$ is the mass of the single fermion $F_j$ assumed to run along
the loop in Fig.\ref{trif-fig} and $N_F$ is a counting factor
for colour and/or hypercolour. \par

In the opposite  \underline{ heavy fermion} $F_j$ limit
where $M^2_j \gg s,~m^2_Z$, we get
\bqa
&& f^{Z}_5 \simeq \frac{e^2 g_{aj}N_F }
{960\pi^2s^3_Wc^3_W} \left({m^2_Z\over M^2_j}\right )
\Big [ 5 g_{vj}^2 +g^2_{aj} +~
\frac{(2s +3\mzd)(7g_{vj}^2 +g^2_{aj})}{21 M^2_j} \Big ]
 ~~ , \label{f5Zmodel}\\
&& h^Z_3\simeq -f^{\gamma}_5 \simeq
-~ {e^2 Q_j g_{vj} g_{aj}N_F\over96\pi^2s^2_Wc^2_W}
\left ({m^2_Z\over M^2_j}\right )
\Big [1 +~ \frac{2(s+\mzd)}{15 M_j^2} \Big ]
 ~~, \label{h3Zmodel} \\
&& h^{\gamma}_3
\simeq -~ {e^2 Q^2_j g_{aj}N_F \over48\pi^2s_Wc_W}
\left ({m^2_Z\over M^2_j}\right )
\Big [1 +~ \frac{2s+\mzd}{15M_j^2} \Big ]
 ~ ~ , \label{h3gmodel}
\eqa
which is the  situation applying to NP contributions characterized
by high scale $ \Lambda_{NP} \simeq M_j $. \par

In both these limits, the   relation
\bq
h^Z_3 \simeq -f^{\gamma}_5 ~~ \label{h3Z-f5g}
\eq
holds  \underline{independently}  of the fermion
couplings. Nevertheless,  this relation cannot be exact.
Indeed, as one can see from the expressions of the Feynman
integrals (\ref{h3Z},~\ref{f5g}), the approximate equality
(\ref{h3Z-f5g}) is
violated by the  different
  threshold effects  associated to the $Z\gamma $ and $ZZ$ final
pairs,  in the  triangular graph in Fig.\ref{trif-fig}.
It turns out though that these threshold effects are rather small
and moreover, they rapidly diminish as soon as $s$ goes away from
threshold. In this respect, we have checked that in a
high $M_j$ expansion,
the violation of the equality (\ref{h3Z-f5g}) is very tiny and
of the order  of  $m^6_Z/M^6_j$. The overall conclusion is  therefore,
   that (\ref{h3Z-f5g}) is
approximately correct for most of the range of the
 $s$ and $M_j$ values. This
is also shown by the  numerical applications  presented in Section
4 below.\par

We next mention  the fermion contribution to the
anapole $ZWW$ and $\gamma WW$ couplings, since they are of the
same nature as the above neutral couplings.
Indeed, when one computes the
triangle loop with a doublet of fermions $F,~F'$ of masses
$M_F,~M_{F'}$, one obtains the result  given  in
(\ref{zZFFp}-\ref{IF}) in Appendix B.
Using then (\ref{zZFDq}- \ref{IFD}) for  the case of standard couplings
and  a degenerate fermion iso-doublet pair satisfying
$M_F=M_{F'} \gg m_W$, that result simplifies to
\bq
z_Z \simeq -{s_W\over c_W} z_{\gamma}=
N_F{e^2\over1152\pi^2s_Wc_W}\left ({m^2_W\over M^2_F}\right )
\left (1+{4s+2m^2_W\over15 M^2_F}\right ) ~ ,
\label{zVgmodelq}
\eq
\noindent
for quarks, and
\bq
z_Z \simeq -{s_W\over c_W} z_{\gamma}=
-N_F{e^2\over384\pi^2s_Wc_W}\left ({m^2_W\over M^2_F}\right )
\left (1+{4s+2m^2_W\over15 M^2_F}\right ) ~ ,
\label{zVgmodell}
\eq
\noindent
for leptons.\par

The numerical predictions for the fermion contributions to
$f_5^V$ and $h_3^V$, are strongly affected by the way the
anomalies are  (presumably) cancelled in the complete theory.
Consequently, in the final part of this section we discuss
their effect. Such anomalies are generated  by triangular fermion
loop diagrams. In SM they are cancelled whenever
 one or more complete families of leptons and quarks are
considered. In the  MSSM the total chargino or
neutralino contributions are separately anomaly free.
However, since these anomalies are proportional to the quantities
\bqa
\sum_j g_{vj} g_{aj} Q_j,  ~~ & ~~  \sum_j g_{aj} Q^2_j ,
~~   &  \sum_j Q_j ~~ ,    \nonumber
\eqa
and independent of the masses of the fermions running along the
loop, their cancellation may  involve fermions with very large
mass differences. Consequently three distinct possibilities arise.
Either all participating fermions are almost degenerate at
scale $\Lambda_{NP}$;
or they have mass differences of the electroweak size; or
finally  certain fermions are much lighter
than others, the contributions of the heavier ones
to the neutral couplings being then negligible.\par

These different situations lead  to very different predictions
for the size of the neutral couplings. In the first case,
a complete family of exactly degenerate heavy fermions
(for example degenerate heavy leptons and quarks with the SM structure)
would lead to the vanishing
of all the NP couplings.
This arises because, in this case, the combination of the heavy fermion
contributions is the same as in the mass independent cancelling
triangle anomaly. This is the
unbroken $SU(2)\times U(1)$ situation.\par

If instead,  one introduces  mass splittings of the electroweak size
(i.e. $\simeq m^2_Z$ )  among the multiplets;
like \eg\@ between the heavy lepton and the heavy quark doublets, then
the resulting couplings are of the order ${m^4_Z/ M^4_F}$,
which means that they are suppressed by an extra
power  $\mzd/M_F^2$ as compared to (\ref{f5Zmodel}-\ref{h3gmodel}).
This case corresponds to a spontaneous broken $SU(2)\times U(1)$
situation. Identifying $M_F$ with $\Lambda_{NP}$, we then indeed get
a contribution of order $1/\Lambda^4_{NP}$, similar to what is
predicted by $dim=8$ gauge-invariant operators.\par

Finally, if a single
heavy fermion (or a partial set of heavy fermions) is much lighter
than all the other fermions in the family, then the couplings are
directly given by the leading terms
in (\ref{f5Zmodel}-\ref{h3gmodel}); \ie\@
just proportional to $({m^2_Z/ M^2_F})$.
This is obviously the most favorable situation
for their observability, but it  would essentially
mean that $SU(2)\times U(1)$ gauge symmetry is strongly broken in
the NP sector. \par

In the next Section we give some quantitative illustrations.

\section{Contributions from Standard Model and Beyond}

\subsection{ The Standard Model contributions}

The SM contribution arises from the three families of
leptons and quarks, and the anomaly cancellation occurs
separately inside each family. To  the couplings  $h^{Z,\gamma}_3$ and
$f^{\gamma}_5$ only the charged fermions contribute; while
$f^{Z}_5$ receives contributions from the neutrinos also.
In Fig.\ref{SM-fig} we have drawn
the SM  contributions to the real and the imaginary parts
of these couplings. The respective magnitudes of the contributions
of the leptons,  the five light quarks, and   the top quark
are also shown. The consequences of the anomaly
cancellation are  clearly reflected in  the behaviour of the predicted
couplings at high energy \ie~ for $s \gg m^2_Z, M^2_F$.
Indeed it can be seen in  Fig.\ref{SM-fig} that
although each fermionic contribution
decreases like $\sim 1/s$ in agreement with eq.(\ref{h3Zs}-\ref{h3gs}),
their sum  decreases like $\sim \ln s/s^2$.\par

The neutral couplings get imaginary parts as soon
as $s >4M^2_F$ or $M^2_Z> 4M^2_F$.  After a spectacular threshold
enhancement though, the imaginary contribution of each fermion
behaves like $(m^2_Z M^2_F/ s^2) \ln(s/M^2_F)$
for $s \gg (M^2_Z,~4M^2_F)$. Therefore, the light
fermion  contribution to the imaginary parts of the couplings
is strongly suppressed, and  only the top quark effect
is visible on Fig.\ref{SM-fig}.\par

Summarizing, the SM neutral couplings are in general complex
but the relative importance of the real and  imaginary parts is
strongly energy dependent. Below the $2m_t$ threshold, the imaginary
part is negligible. Above $2m_t$, real and imaginary parts have a
comparable magnitude; with the imaginary part being somewhat larger.
In Table 1, we collect these SM contributions
at LEP2 ($\sqrt{s}=200~ GeV$) and at a $500~GeV$ Linear
Collider. The results in Fig.\ref{SM-fig} and Table 1 indicate also
the extent to which (\ref{h3Z-f5g}) is approximately correct for
any $s$ and $M_j$ values.

\begin{table}[htb]
\begin{center}
{ Table 1: SM contributions in units of $10^{-4}$ .}\\
\vspace*{0.3cm}
\begin{tabular}{|c|c|c|c|c|}
\hline
\multicolumn{1}{|c|}{$\sqrt{s}$}&
\multicolumn{1}{|c|}{$f^{\gamma}_5$}&
\multicolumn{1}{|c|}{$f^{Z}_5$ }&
\multicolumn{1}{|c|}{$h^{\gamma}_3$}&
\multicolumn{1}{|c|}{$h^{Z}_3$}
\\[0.1cm] \hline
$ 200$ & $2.05-0.15\times10^{-4}~ i $ & $1.85-0.48\times10^{-4}~i $
& $-7.21+0.81\times10^{-2}~i $ & $-2.10+0.40\times10^{-2}~i$\\
\hline
$500 $ & $0.25+0.62~i $ & $0.46+0.53~i $ & $-0.88-1.82~i $
& $-0.26-0.62~i $
\\[0.0cm] \hline
\end{tabular}
\end{center}
\end{table}

\subsection{ The MSSM contributions}

As a first example of new physics effect, we have computed
the additional one loop contributions arising in the Minimal
Supersymmetric Standard Model (MSSM). They are rather simple,
as the only new fermions are charginos and neutralinos.
The two charginos $\chi^{\pm}_{1,2}$ contribute to the four
$h^{Z,\gamma}_3$ and $f^{Z,\gamma}_5$ couplings; while the four
neutralinos $\chi^0_{1-4}$  contribute only to  $f^{Z}_5$. Charginos
couple to the gauge bosons  through both
their gaugino and higgsino components,
whereas neutralinos only contribute through their higgsino
components. The new feature, as compared to SM contributions,
is that there now exist triangle loops with mixed contributions, for
example $(F_1,~F_2,~ F_2)$ in the chargino case. This is caused by the
non-diagonal  $g_{v12},~ g_{a12}$   Z-couplings in (\ref{VheavyF}).
The explicit expressions of these new contributions
are also given in Appendices A and B.
However, because the  non-diagonal Z-couplings are generally weaker
than the diagonal ones (see below),
these mixed contributions turn out to be notably smaller
than the unmixed contributions.\par

Let us first discuss the chargino contributions.
 If $\mu$ and the soft breaking SUSY parameters are taken to be
real, as would be the case if no new CP violation source, beyond
the one contained in the Yukawa couplings exists; then the
 chargino masses are given by \cite{MSSM}
\bq
M^2_{ \chi_{1,2}}={1\over2} \left \{M^2_2+\mu^2+2\mwd \mp
\sqrt{(M^2_2+\mu^2+2\mwd )^2-4[M_2\mu-\mwd \sin(2\beta)]^2}~
\right \} . \label{chimass}
\eq
The photon couplings in (\ref{VheavyF}) are then
fixed by $Q_{\chi_{1,2}}=+1$.\par

For the $Z\chi_1\chi_1$ couplings in (\ref{VheavyF})
we have \cite{MSSM}
\bq
g_{v1}= {3\over2}-2s^2_W+{1\over4}[\cos2\phi_L+\cos2\phi_R]~~ , ~~
g_{a1}=- ~{1\over4}[\cos2\phi_L-\cos2\phi_R] ~ ,
\label{Zch1}
\eq
while for the $Z\chi_2\chi_2$ couplings
\bq
g_{v2}= {3\over2}-2s^2_W-~{1\over4}[\cos2\phi_L+\cos2\phi_R]~~ , ~~
g_{a2}={1\over4}[\cos2\phi_L-\cos2\phi_R] ~ ,
\label{Zch2}
\eq
\noindent
where
\bqa
\sin2\phi_R=-~{2\sqrt{2}m_W(\mu \cos\beta+M_2 \sin\beta)\over D} &,&
\sin2\phi_L=-~{2\sqrt{2}m_W(\mu \sin\beta+M_2 \cos\beta)\over D}\nonumber\\
\cos2\phi_R=-~{M^2_2-\mu^2+2m^2_W \cos2\beta\over D} &,&
\cos2\phi_L=-~{M^2_2-\mu^2-2m^2_W \cos2\beta\over D}~ ,
\label{phiLR}\eqa
\noindent
with
\bq
D=\sqrt{(M^2_2+\mu^2+2m^2_W)^2 -4(M_2\mu-m^2_Wsin2\beta)^2} ~ .
\eq
Finally,  the mixed $Z\chi_1\chi_2$ couplings are
\bqa
&&g_{v12}=
-~{\Sn(M_2)\over4}
[\Sn(M_2\mu-m^2_W\sin2\beta)\sin2\phi_R+\sin2\phi_L] ~ ,
\nonumber\\
&&g_{a12}=-~{\Sn(M_2)\over4}
[\Sn(M_2\mu-m^2_W\sin2\beta)\sin2\phi_R -\sin2\phi_L] ~ ,
\label{Zch12}
\eqa
where $\Sn(x)$  means sign of $x$.
As expected, the anomaly cancellation in the chargino sector
 arises when one sums the
mixed and unmixed contributions of the two charginos.  \par

For the numerical applications we used the sets of parameters
at the electrowaek scale presented in Table 2, \cite{charinput}.
%
%
%
\begin{table}[htb]
\begin{center}
{ Table 2: Sets of MSSM chargino parameters at the electroweak scale.}\\
\vspace*{0.3cm}
\begin{tabular}{|c|c|c|c|c|c|}
\hline
\multicolumn{1}{|c|}{Set}&
\multicolumn{1}{|c|}{$M_2$}&
\multicolumn{1}{|c|}{$\mu$ }&
\multicolumn{1}{|c|}{$\tan\beta$}&
\multicolumn{1}{|c|}{$M_{\chi_1^+}$}&
\multicolumn{1}{|c|}{$M_{\chi_2^+}$}
\\[0.1cm] \hline
$ (1)$ & $81 $ & $-215 $
& $2$ & $94.71$ &$237.94$\\ \hline
$ (2)$ & $215 $ & $-81 $
& $2$ & $94.71$ &$237.94$\\ \hline
$ (3)$ & $120 $ & $300 $
& $2.5$ & $96.13$ &$328.57$\\ \hline
$ (4)$ & $200 $ & $800 $
& $4$ & $193.95$ &$809.43$\\ \hline
$ (5)$ & $152 $ & $316 $
& $3$ & $127.89$ &$345.55$\\ \hline
$ (6)$ & $150 $ & $263 $
& $30$ & $132.34$ &$294.88$
\\[0.0cm] \hline
\end{tabular}
\end{center}
\end{table}

 The main feature  of the data in Table 2  is that almost always
one of the chargino is considerably  lighter
than the other. Thus, at  energies around $2 M_{\chi_1}$ but
considerably below $2M_{\chi_2}$,
the dominant contribution comes from the lighter chargino, while the
heavier one gives a
weaker contribution of opposite sign. At  energies
higher than both chargino thresholds, these two contributions tend to
cancel; the cancellation being a relic of the anomaly
cancellation in the chargino sector.\par

More quantitatively, as the chargino couplings are of electroweak size,
when one chargino
is light enough (around $100~GeV$ or less), then their
contribution is
similar to the SM ones, for what concerns the real parts of the
couplings. Similarly  to the top quark SM case, the imaginary
parts get  threshold enhancements
just above $2M_{\chi}$, where their magnitude is
comparable to that of the real part. These features can be
seen in Figs.\ref{Set5-fig}, \ref{Set6-fig}
where we have illustrated the cases of the Sets 5 and 6.
The results for the Sets 1, 2, 3 are quite similar; although
 it may remarked that the chargino contribution, particularly
to   $f_5^Z$ or $f_5^\gamma$, is somewhat more pronounced
there. In Set 4 the  chargino contribution
is very small, because there, the lightest chargino is
relatively  heavy,  and the  Z-axial couplings are very small.
The later is due to  $|\mu| \gg |M_2|$ in this model,
which forces  $\phi_L, ~ \phi_R$  to have similar values
and thereby the Z-axial couplings
to have very low values.
The cases of Sets 5 and 6 are identical to the
low and  high $\tan\beta$ scenarios suggested in
\cite{charbest} as a benchmark
for SUSY studies.  In Table 3
we  give the precise predictions for
these  sets at $\sqrt{s}=200,~500~ GeV$. \par

\vspace*{0.3cm}
\begin{table}[htb]
\begin{center}
{ Table 3: Chargino contributions for Set (5,6)
in units of $10^{-4}$.}\\
\vspace*{0.3cm}
\begin{tabular}{|c|c|c|c|c|c|}
\hline
\multicolumn{1}{|c|}{Set}&
\multicolumn{1}{|c|}{$\sqrt{s}$}&
\multicolumn{1}{|c|}{$f^{\gamma}_5$}&
\multicolumn{1}{|c|}{$f^{Z}_5$ }&
\multicolumn{1}{|c|}{$h^{\gamma}_3$}&
\multicolumn{1}{|c|}{$h^{Z}_3$}
\\[0.1cm] \hline
 (5) & $ 200 ~GeV $ & $0.29$ & $0.49 $
& $-0.19 $ & $-0.31$\\
\hline
 (5) & $500~GeV $ & $-0.17+ 0.13\ i $ & $-0.39 + 0.11\ i $
 & $0.11-0.10\ i $ & $0.26-0.09\ i $ \\
\hline\hline
 (6) & $ 200 ~GeV $ & $0.43$ & $0.70 $
& $-0.29 $ & $-0.46$\\
\hline
 (6) & $500~GeV $ & $-0.34+ 0.25\ i $ & $-0.57 + 0.15\ i $
 & $0.22-0.20\ i $ & $0.39-0.12\ i $
\\[0.0cm] \hline
\end{tabular}
\end{center}
\end{table}

\begin{figure}[p]
\vspace*{-4cm}
\[
\epsfig{file=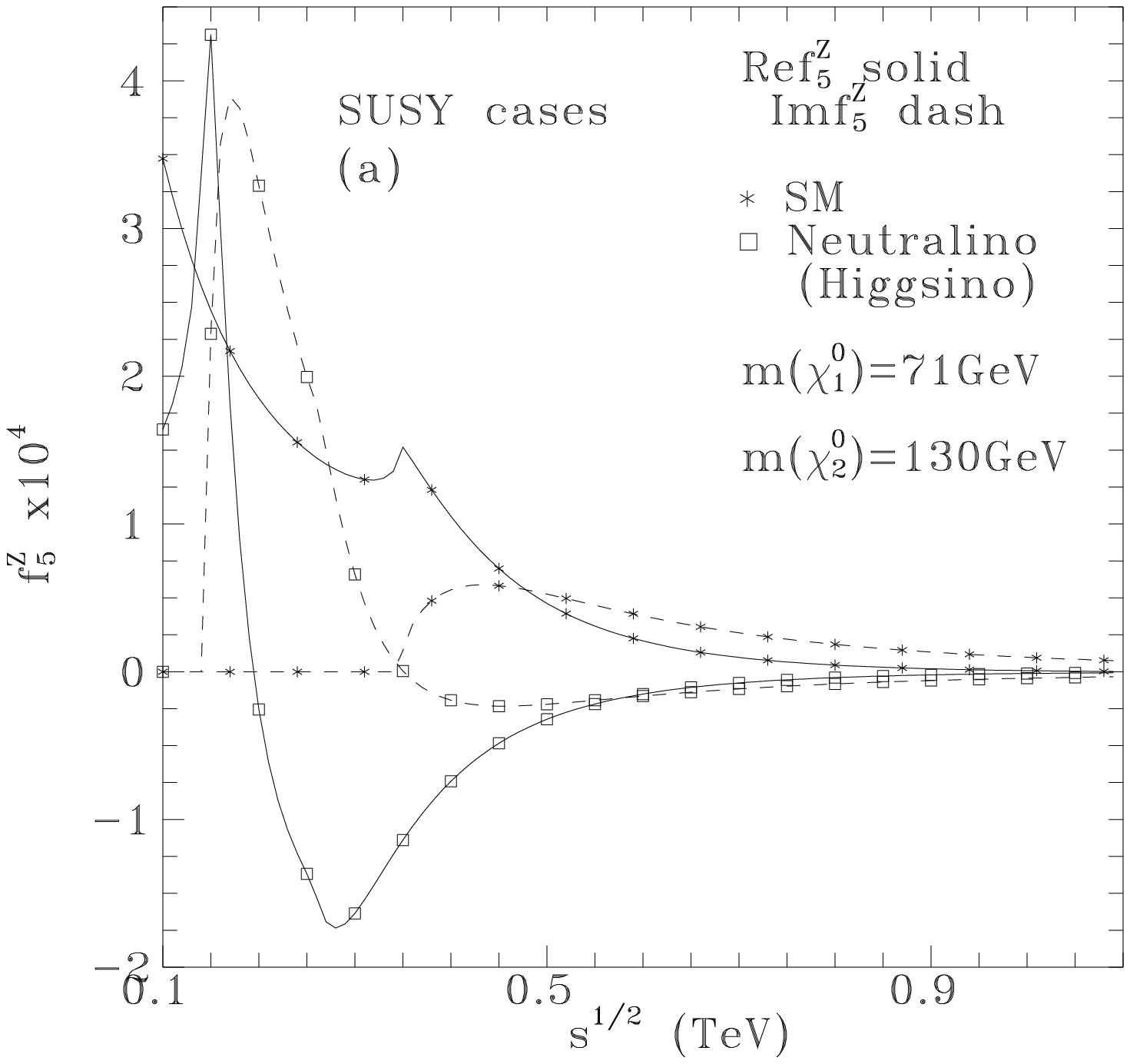,height=7.5cm}\hspace{0.5cm}
\epsfig{file=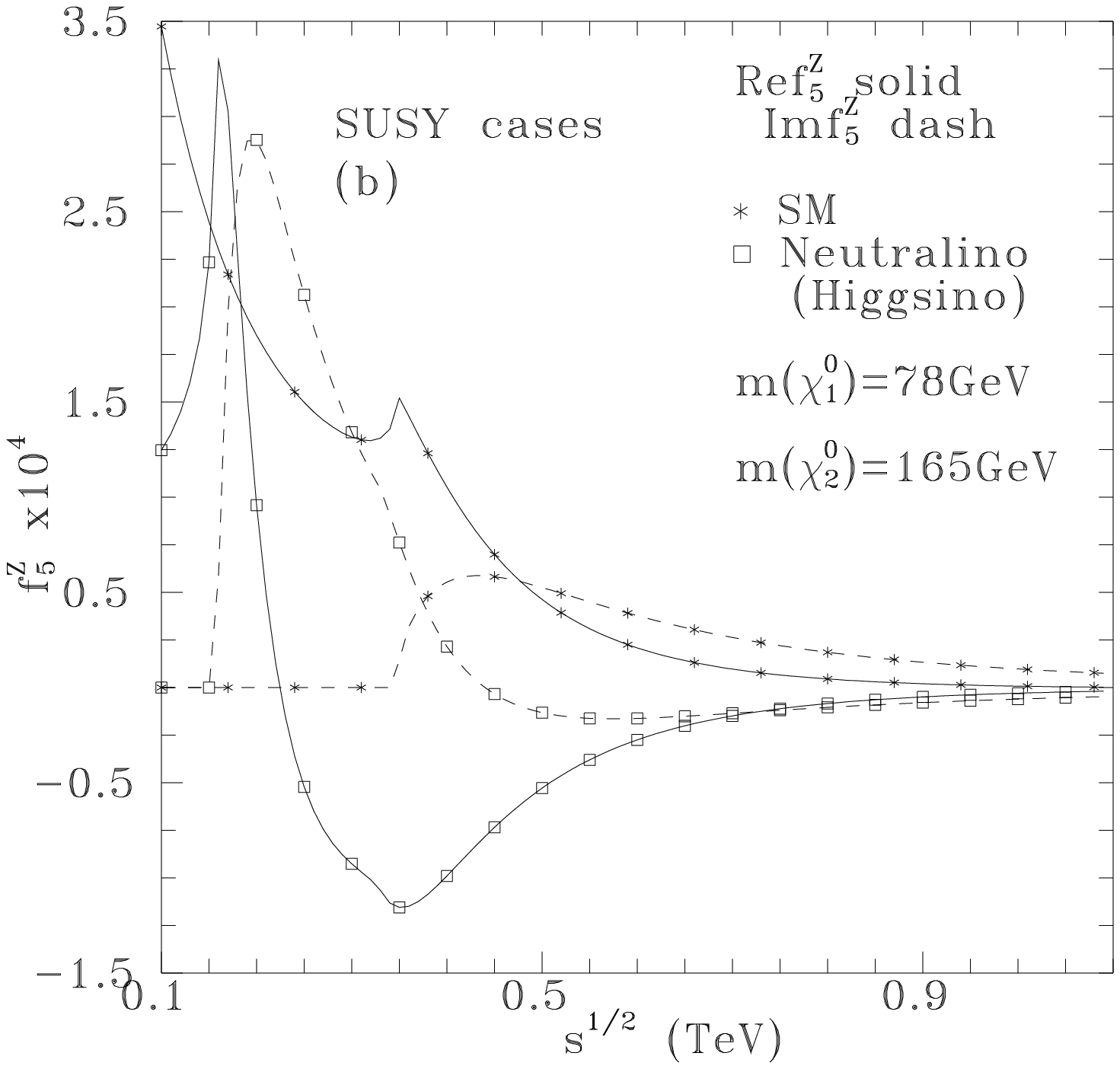,height=7.5cm}
\]
\vspace*{0.5cm}
\[
\epsfig{file=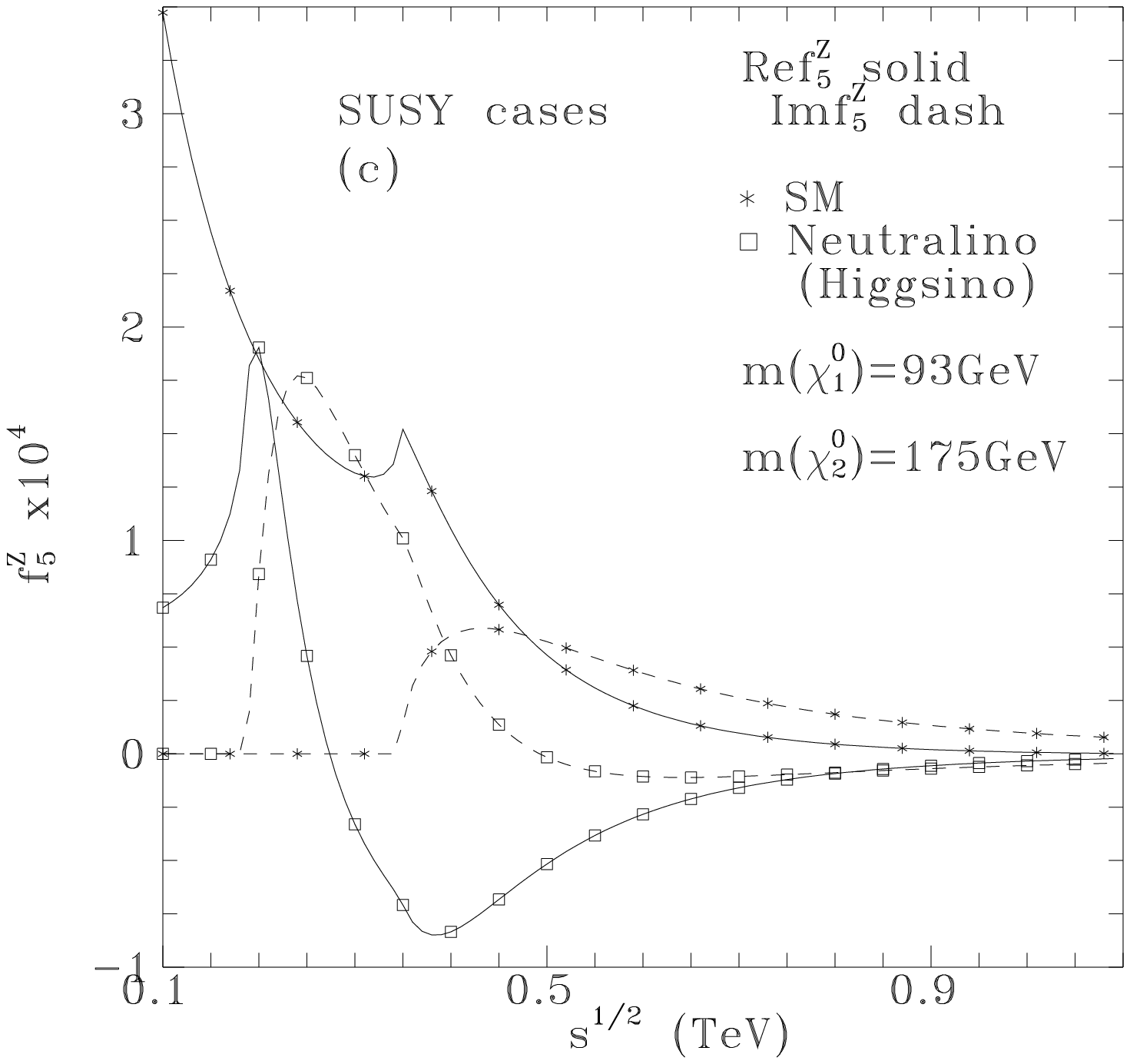,height=7.5cm}\hspace{0.5cm}
\epsfig{file=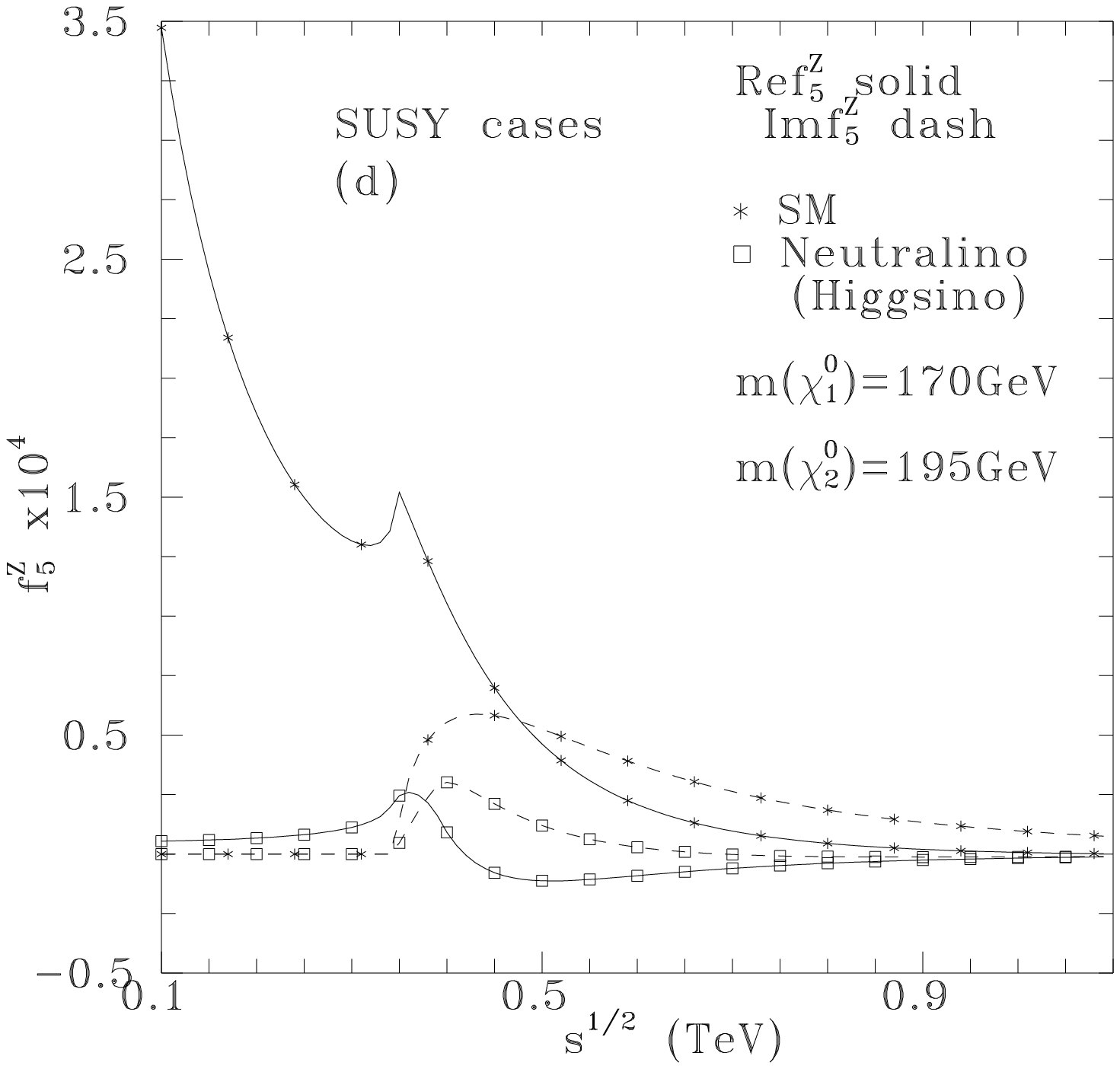,height=7.5cm}
\]
\vspace*{0.5cm}
\caption[1]{Contribution  to $f_5^Z$ from two higgsino-like
neutralinos  with  the indicated masses.}
\label{neutralino-fig}
\end{figure}

In addition there is a neutralino contribution to $f^Z_5$. It only
arises from the $Z$ couplings to the Higgsino components. In order to
have an estimate of the largest possible neutralino effect, we have
taken the case in which the lightest neutralinos are of Higgsino type.
As the contribution to $f^Z_5$ is proportional to the cubic power of the
$Z$-neutralino coupling, it would become rapidly negligible if
the neutralinos were not dominantly Higgsino-like.
So we have taken pure axial Z-Higgsino couplings
\bq
g_{v\chi^0_1}=g_{v\chi^0_2}= 0~~, ~~
g_{a\chi^0_1}=-g_{a\chi^0_2}=1 ~ ,
\label{Zn}
\eq
and only considered the contribution from two neutralinos.
The results for the  masses (in GeV) \cite{neutinput}
\bq
(M_{\chi^0_1},~M_{\chi^0_2})=~ (71,130),~~ (78,165),~~ (93,165),
~~ (170,195) ~~, \label{neutralino-masses}
\eq
are shown in Fig.\ref{neutralino-fig}. They are somewhat smaller
than those  due to charginos for comparable masses, this
being due to the difference in the $Z$ couplings. Remarkable threshold
effects nevertheless appear when one neutralino is light enough.
The largest effects are
obtained for the couple (71,130)GeV.\par

For this couple, at 200 and at 500 GeV, the
contributions to $f^Z_5$ in units of $10^{-4}$, are
\bq
-0.26+3.3~i ~~~ ~~ \mbox{ and} ~~~~   -0.32-0.22~i  \nonumber
\eq
respectively; \ie~  comparable to the SM and to chargino
contributions. The same type of comments about the behaviour of the
real and of the imaginary parts can be made.\par
Let us finally note that, if there is nearly degeneracy of the lightest
chargino and neutralino, their cumulative effect in $f^Z_5$ can be
occasionally important and increase the SM contribution by a
factor of 2 to 3.

\subsection{ Other NP possibilities.}

The above study of the MSSM contributions gives already a good feeling
of what can arise from any other perturbative NP contributions
at one loop.
We now want to extend the discussion by considering quantitatively,
in a model-independent way,
the contribution of a new fermion $F$. Such new fermionic
states arise in many extensions of the SM, like in GUTS or in TC
models.\par

We take couplings of electroweak size, which means typical values
of the order of $g_{vF} =g_{aF}={1\over2}$, $Q_F=1$, and
keep the colour-hypercolour factor $N_F=1$.
In Fig.\ref{NPfermion-fig} we show how the resulting couplings
depend on the fermion mass $M_F$, at fixed energies
$\sqrt{s}=200,~500~GeV$. The cusps in the real
parts and the peaks in the imaginary parts that occur around
$M_F=\sqrt{s}/2$, are clearly visible there.
They have the same structure as in the previous SM
and MSSM cases and with the chosen  couplings,  their magnitude
is similar,  \ie~ a few $10^{-4}$.\par

However for
$M_F \gg \sqrt{s}/2$, which is the
case that we now want to discuss, the $M_F$ dependence is smooth
and  the resulting neutral couplings are
\underline{purely real}. Thus in this limit, they can be well
approximated by the energy-independent empirical formulae
\bqa
h^{\gamma}_3& =& -0.02\times 10^{-4}
\left ({1~TeV\over M_F}\right )^2 ~ , \label{NP-h3g} \\
h^{Z}_3& =& - f^{\gamma}_5=-0.01\times 10^{-4}
\left ({1~TeV\over M_F}\right )^2
~ ,  \label{NP-h3Z-f5g} \\
f^{Z}_5 & = & 0.009\times 10^{-4}
\left ({1~TeV\over M_F}\right )^2 ~ ; \label{NP-f5Z}
\eqa
\noindent
(compare with eq.(\ref{f5Zmodel}-\ref{h3gmodel})). \par

As they stand, these results would describe the most favorable NP
case in which a single fermion is lighter than the other ones.
With the chosen electroweak couplings and taking $N_F=1$,
such a contribution becomes nevertheless quickly unobservable
when the mass $M_F$ reaches the level of a few hundred of GeV.
It could only be sizeable
if the heavy fermions have enhanced couplings to the photon or to
the $Z$ (see Section 5), or if
the  colour-hypercolour factor $N_F$ is large.
Consequently, it is unlikely that such a new contribution
(even without the
depressing effect of the anomaly cancellation),
can appreciably modify the SM prediction.

\section{Higher orders and Non Perturbative effects}

At the one-loop level we have
exploited so far, only the fermionic triangle diagram can contribute
to the generation of neutral gauge couplings. In such a case, the CP
conserving $h^{\gamma, Z}_4$ couplings are never generated.
This is a direct consequence of the symmetries of
the fermionic trace of the triangular diagram and of Shouten's relation.
It is therefore interesting to examine if there is any other  way to
generate such $h^{\gamma, Z}_4$ couplings and enhance its magnitude;
compare eq.(\ref{hZgamma}).\\

\begin{figure}[p]
\vspace*{-2cm}
\[
\epsfig{file=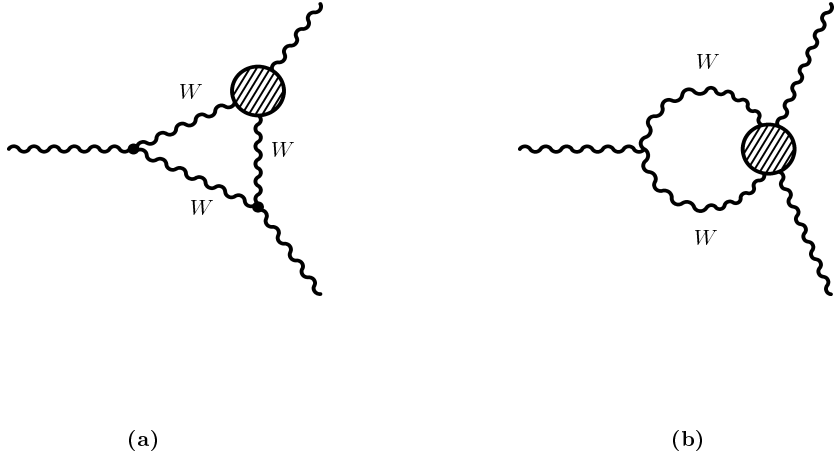,height=6cm,width=11cm}
\]
\vspace*{0cm}
\caption[3]{Additional bosonic contribution
to the fermionic triangle. The $W$ lines represent both
$W^{\pm}$ and Goldstone $\Phi^{\pm}$ contributions.
(a) Bosonic triangle and fermionic triangle
(represented by the hatched blob).
(b) Bosonic bubble and fermionic box
(represented by the hatched blob)}
\label{triWf-fig}
\end{figure}

\noindent
\underline{Perturbatively}, this may happen at a higher-loop level.
We have thus explored diagrams of the form of
Fig.\ref{triWf-fig}a, where the hatched blob denotes a fermion 1-loop
diagram generating an anapole $ZWW$ coupling.
We have  found that such an anapole $ZWW$-vertex,
inside the W-loop of Fig.\ref{triWf-fig}a,  generates
an $h^{\gamma, Z}_4$ coupling of the size
\bq
h^{\gamma, Z}_4 \sim {\alpha\over4\pi}h^{\gamma, Z}_3 ~~;
\label{h4V}
\eq
\ie~ the size of a typical electroweak correction to the NP
prediction to $h^{\gamma, Z}_3$. Of course the result (\ref{h4V})
 should only be considered as a rough order of magnitude
expectation, since a complete
model prediction would require
the computation of other diagrams appearing at the same 2-loop
order, like those depicted in Fig.\ref{triWf-fig}b.\\

\begin{figure}[htb]
\hspace{-1cm}
\[
\epsfig{file=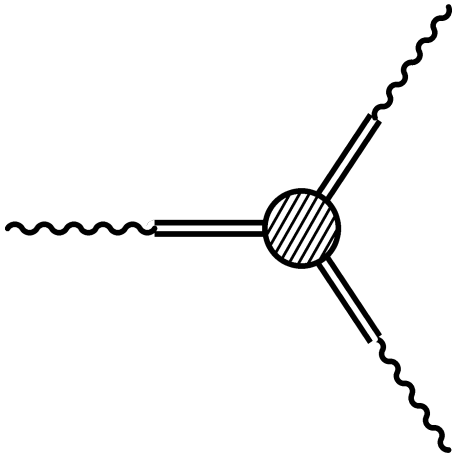,height=6cm,width=6cm}
\]
\vspace*{1cm}
\caption[2]{Contributions from strongly interacting Vector mesons.
Double lines denote the heavy vector bosons $(\V, ~ \A)$, while the
wavy lines describe $Z,$ or $\gamma$.}
\label{had-fig}
\end{figure}

\noindent
It is conceivable that \underline{non-perturbative} effects
could enhance the above neutral gauge boson couplings.
In this respect we may consider  strong vector
$\V$ ($\rho$-like) and axial $\A$ ($A_1$-like) resonances
coupled to the photon and $Z$, like in TC models \cite{TC},
 through  the junctions
\bq
eg_{\V\gamma}=eF_{\V} M_{\V} \ \ , \ \
eg_{\V Z}=e{(1-2s^2_W)\over2s_Wc_W}F_{\V} M_{\V} \ \ , \ \
eg_{\A Z}={e\over2s_Wc_W}F_{\A} M_{\A} ~~ ,
\label{VDM}
\eq
where we  expect that in the strong coupling regime \cite{TC}
\bq
{F_{\V,\A} \over M_{\V,\A}}\simeq O({1\over \sqrt{2\pi}}) ~~ ,
\eq
should hold.
New strong interactions can generate non perturbative
couplings among the $\V$ and $\A$ vector bosons;
like \eg~ those expected  in the Vector Dominance Model (VDM)
of  hadron physics
 for the $\rho$ and $A_1$ vector mesons  \cite{Walsh}.
Such couplings, depicted by the central bubble
in Fig.\ref{had-fig}, could have the same Lorentz decomposition
as in eq.(\ref{hZgamma}), but with  strengths determined by
\bqa
 & h^S_3/\Lambda^2_{NP}~~~ & , ~ ~~~h^S_4/\Lambda^4_{NP} ~~,
 \nonumber
\eqa
where $h^S_{3,4} \sim \sqrt{4\pi}$.
By multiplying these strengths by the junctions
given in eq.(\ref{VDM}) and using
the corresponding $\V$ or $\A$ propagators according to
Fig.\ref{had-fig}, one then obtains  the corresponding
predictions for neutral gauge couplings. For
$s,~m^2_Z \ll M^2_{\V,\A}\simeq\Lambda^2_{NP}$ we then get
\bq
h^{Z,\gamma}_4\simeq {m^2_Z\over M^2_{\V,\A}} h^{Z,\gamma}_3
\simeq\alpha \left ({m^2_Z\over M^2_{\V,\A}}\right )^2 ~~ .
\eq
For vector meson masses $M_{\V,\A}$ not too far in the TeV range,
these values may be somewhat higher than those predicted by
the previous perturbative computations.

\section{Concluding Remarks}

We have studied various ways to generate neutral triple gauge boson
couplings among the  photon and $Z$. Since these couplings are
actually form factors involving at least one  off-shell vector
boson, they  depend  on the corresponding  energy-squared variable $s$.
The simplest way to generate them is through a fermionic
triangle loop, involving fermions   with arbitrary vector and axial
gauge couplings. Such diagrams only generate CP-conserving
couplings satisfying
\bq
h^Z_3\simeq -f^{\gamma}_5 ~~ \ \ \ , \ \ \ \
  h^Z_4\equiv h^{\gamma}_4\equiv 0 ~~ ,
\label{rel}
\eq
for almost any  $s$ and fermion mass.  \par

We have then studied the high energy  behaviour of these couplings
at a fixed fermion  mass; as well a  the
high fermion  mass limit at current energies. This last case allows us to
illustrate different possible situations, that depend on the
NP mass spectrum and on the way the anomaly cancellation takes place.
The most favorable situation arises whenever
one of the states needed to cancel the  anomaly  is much lighter
that the rest.\par

\begin{figure}[p]
\vspace*{-4cm}
\[
\epsfig{file=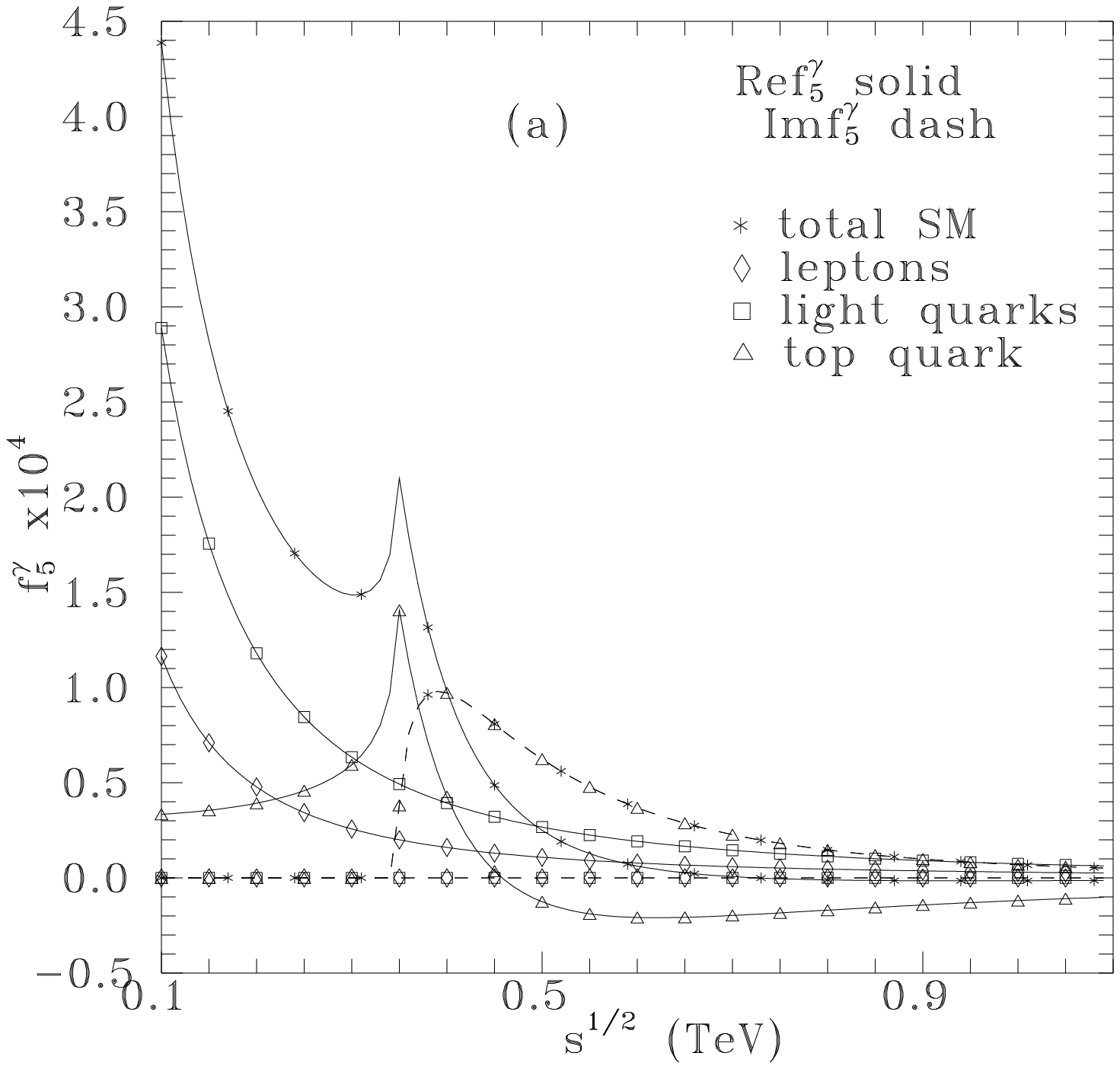,height=7.5cm}\hspace{0.5cm}
\epsfig{file=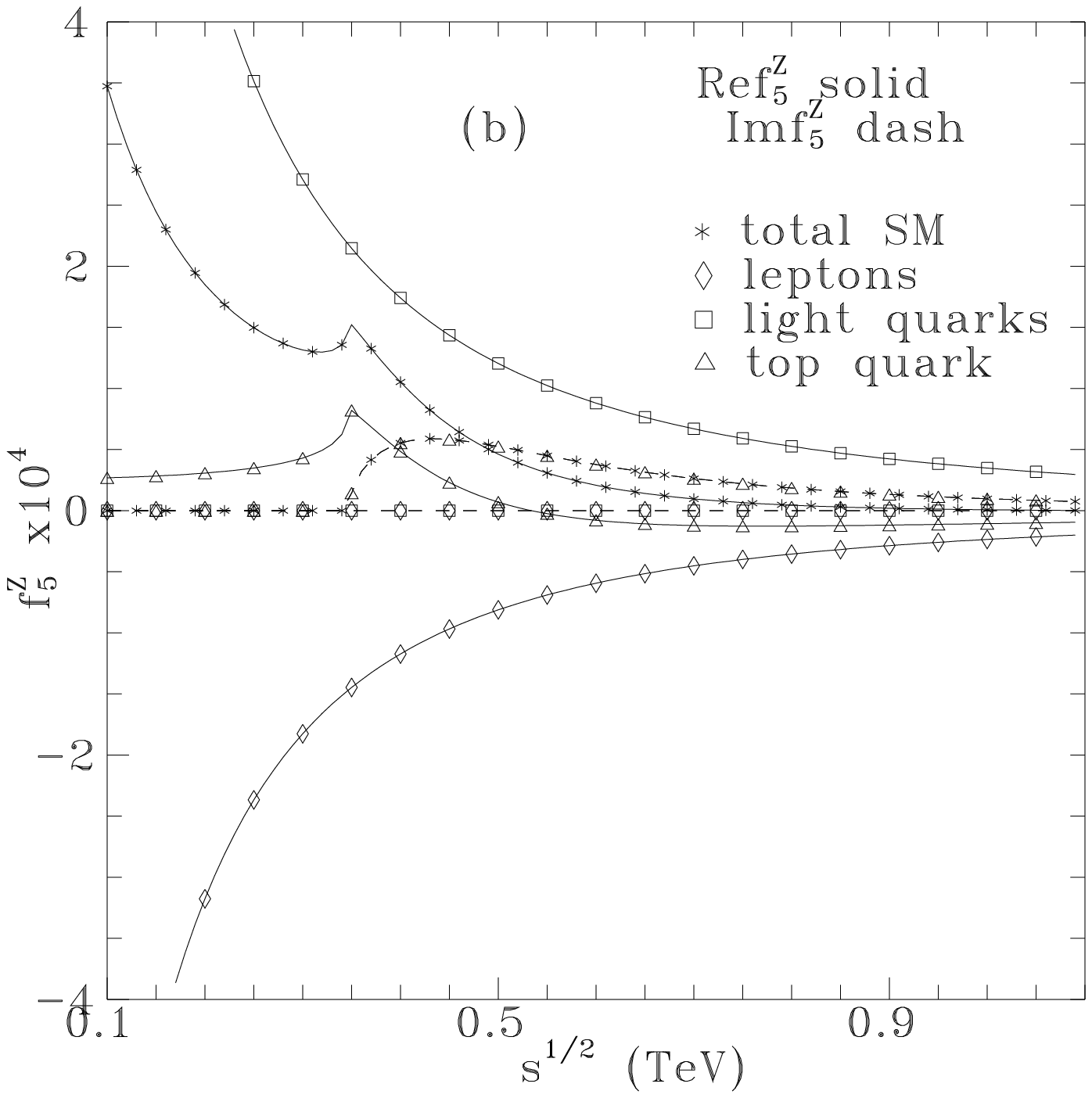,height=7.5cm}
\]
\vspace*{0.5cm}
\[
\epsfig{file=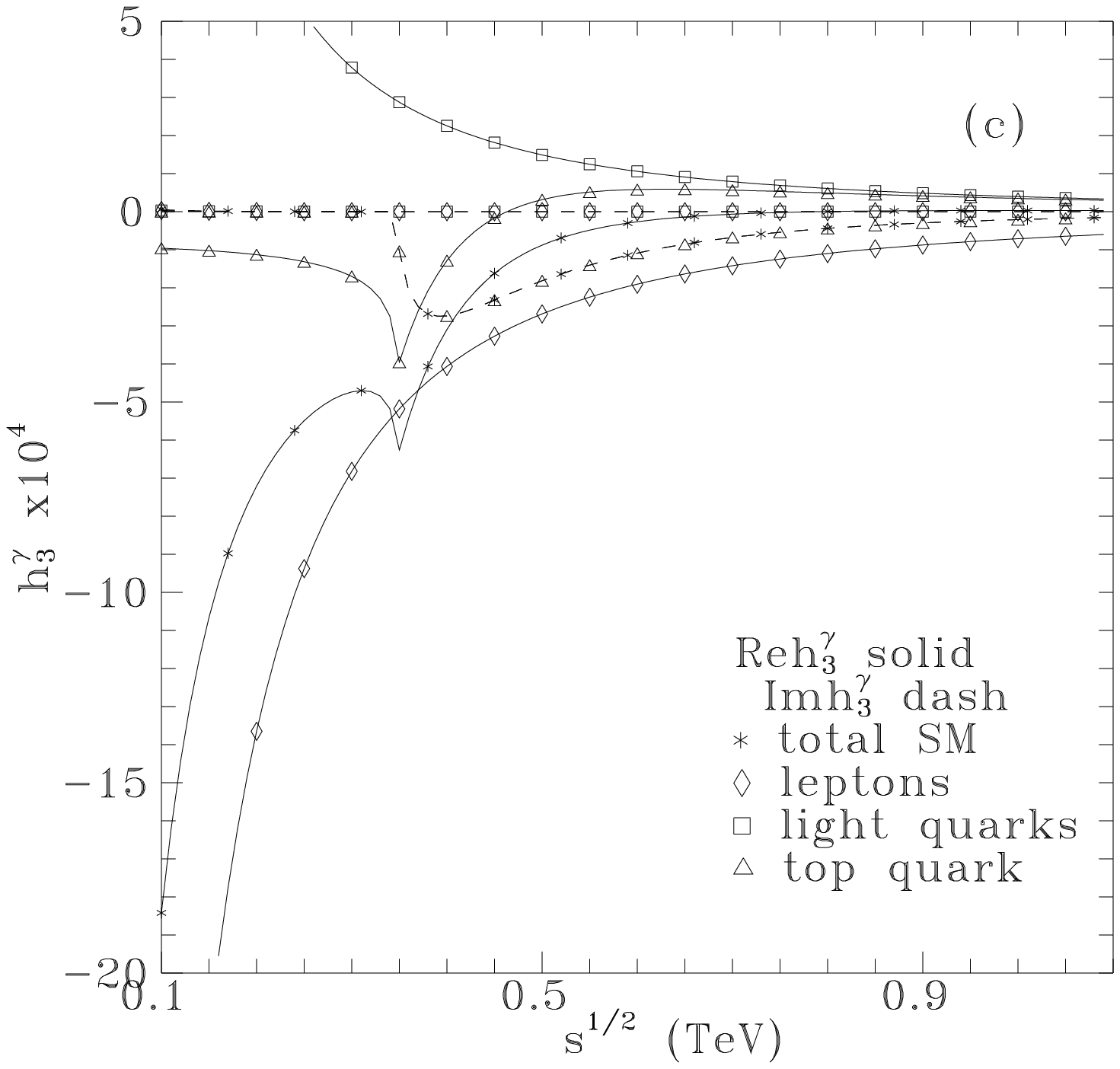,height=7.5cm}\hspace{0.5cm}
\epsfig{file=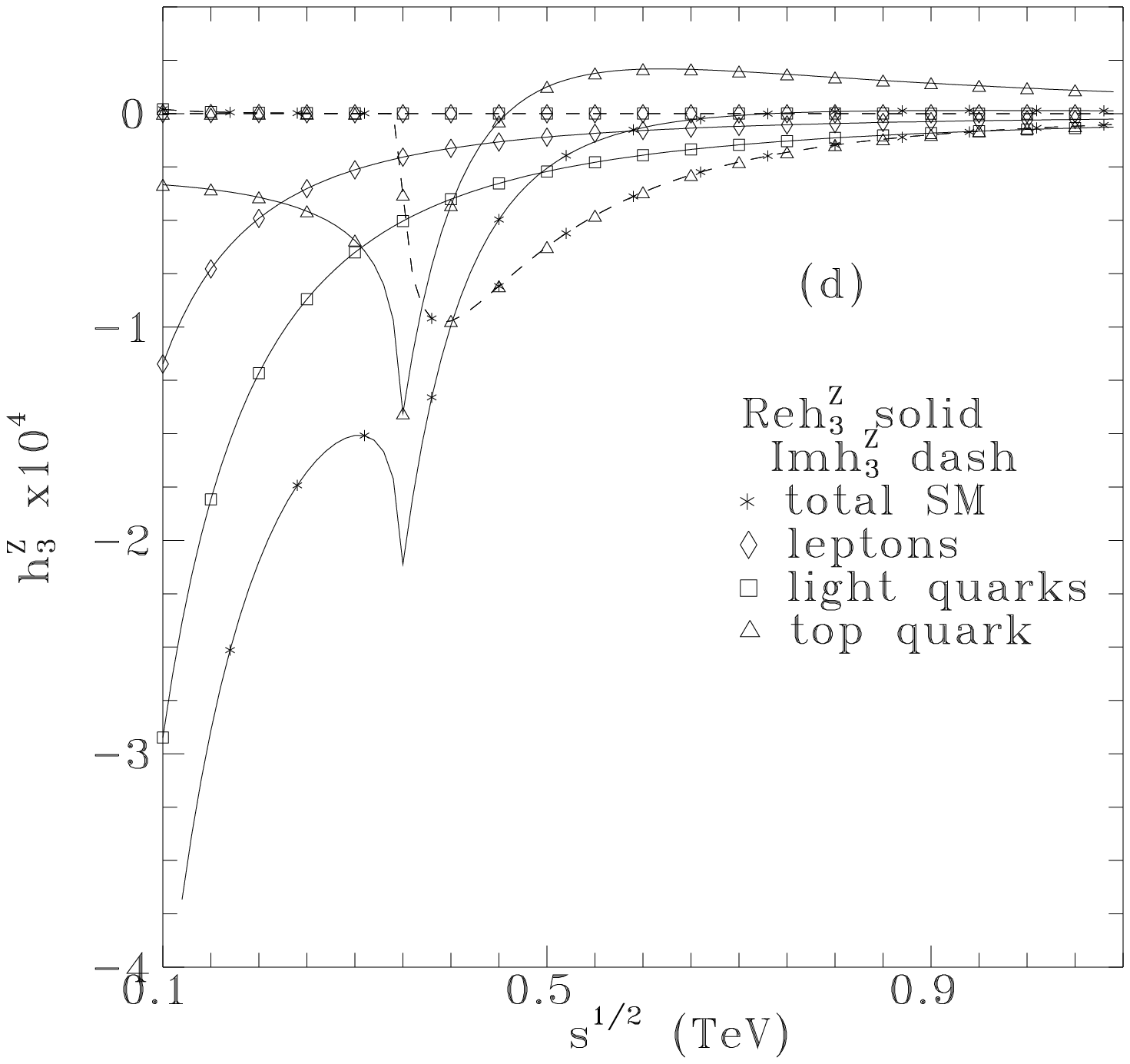,height=7.5cm}
\]
\vspace*{0.5cm}
\caption[1]{The SM  contributions to the neutral gauge boson
self interactions. The separate contributions from the SM leptons,
the top quark and the other quarks are also given.}
\label{SM-fig}
\end{figure}

To acquire a feeling of the expected magnitudes we have presented
in Fig.\ref{SM-fig} the SM prediction for these coupling as
a function of the energy $\sqrt{s}$. These SM predictions are found to be
at the level of a few $10^{-4}$ for LEP (200GeV), while for
LC(500GeV) they reduce to a few $10^{-5}$; see Table 1.
Just above the $2m_{t}$ threshold, an imaginary part  of
the order of  $10^{-4}$ appears due to top quark
contribution. According to
a previous study \cite{neut}, such values
should be unobservable at LEP2, but marginally observable at a high
luminosity LC \cite{LC}.\par

\begin{figure}[p]
\vspace*{-4cm}
\[
\epsfig{file=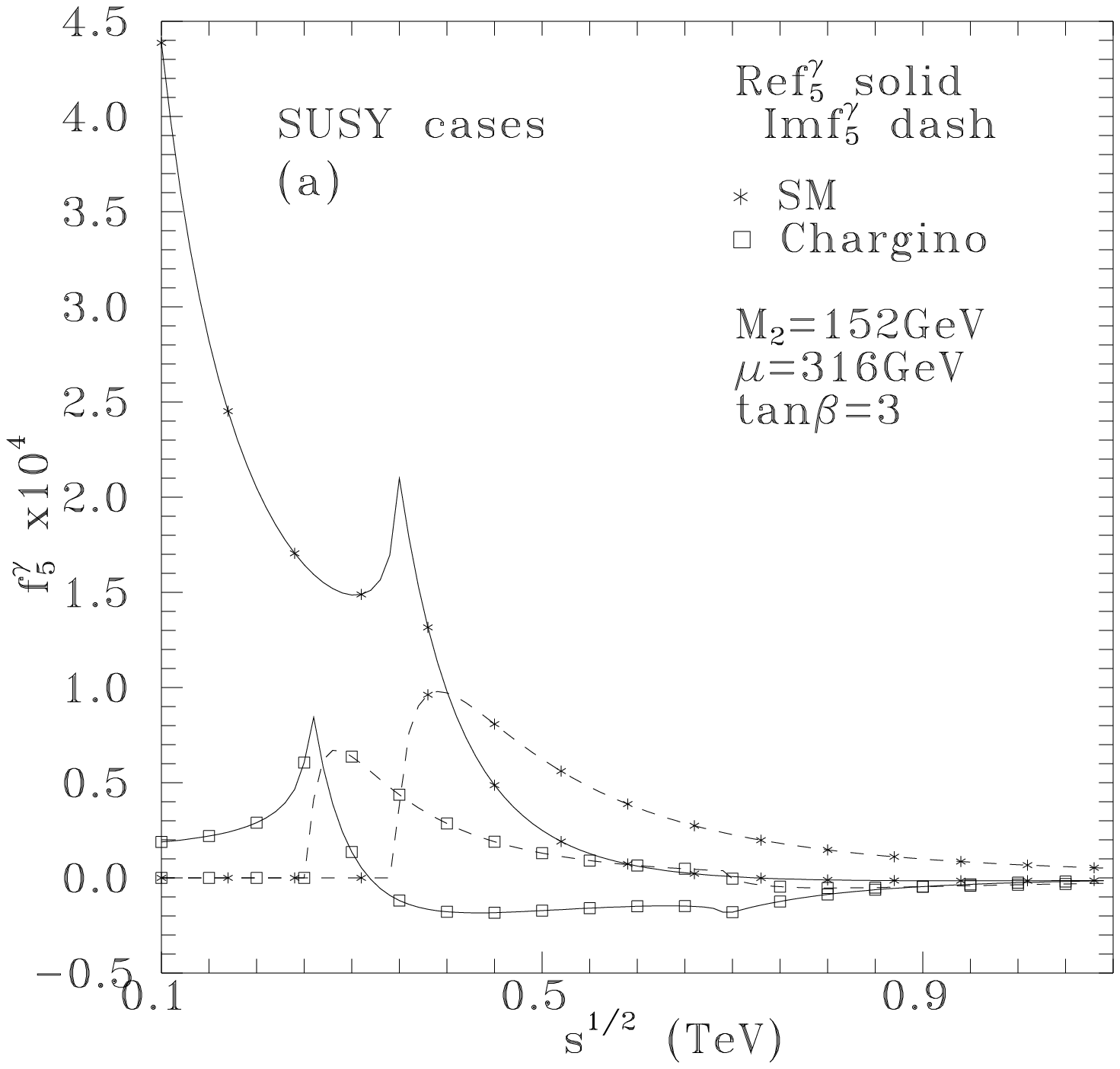,height=7.5cm}\hspace{0.5cm}
\epsfig{file=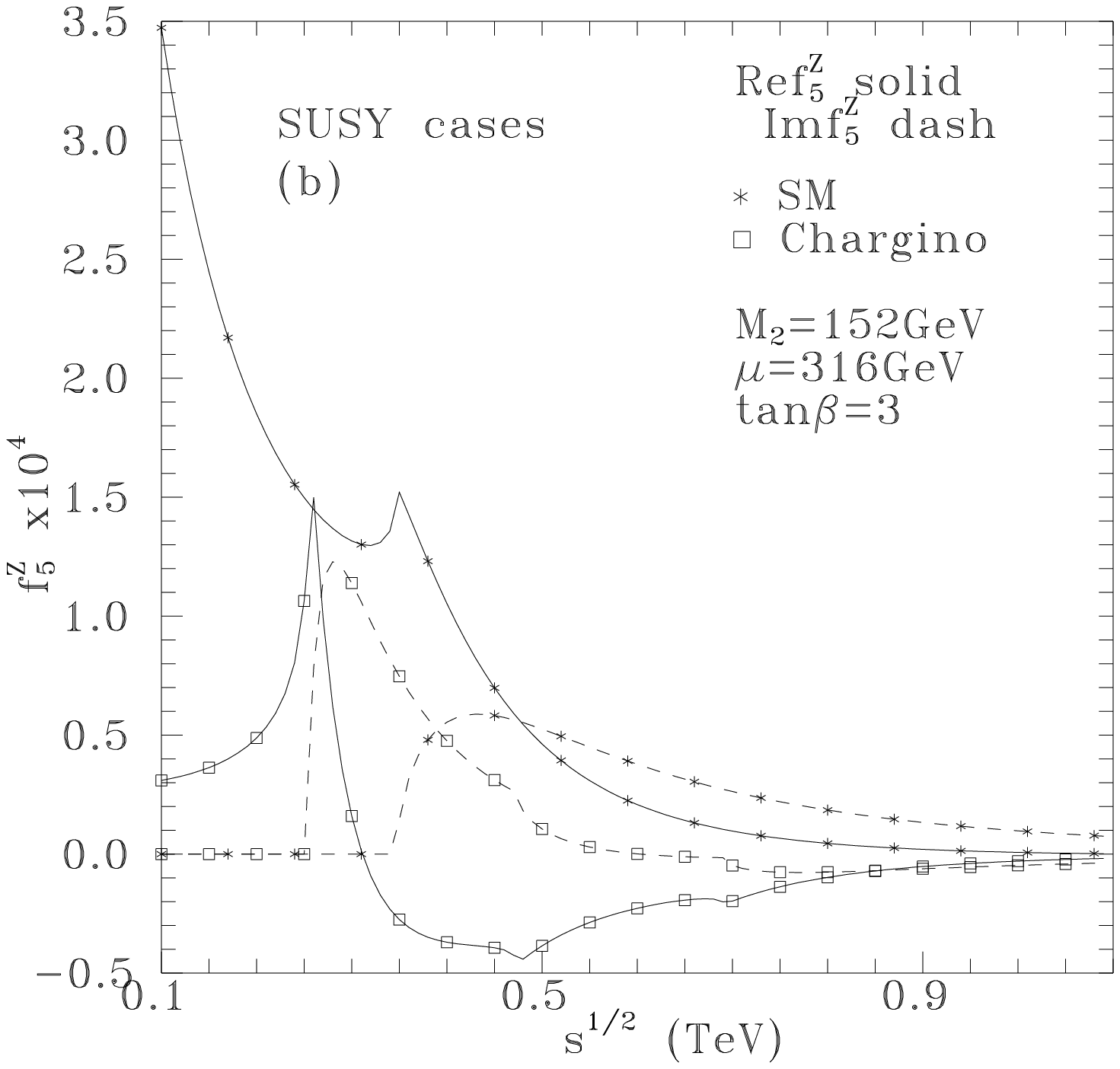,height=7.5cm}
\]
\vspace*{0.5cm}
\[
\epsfig{file=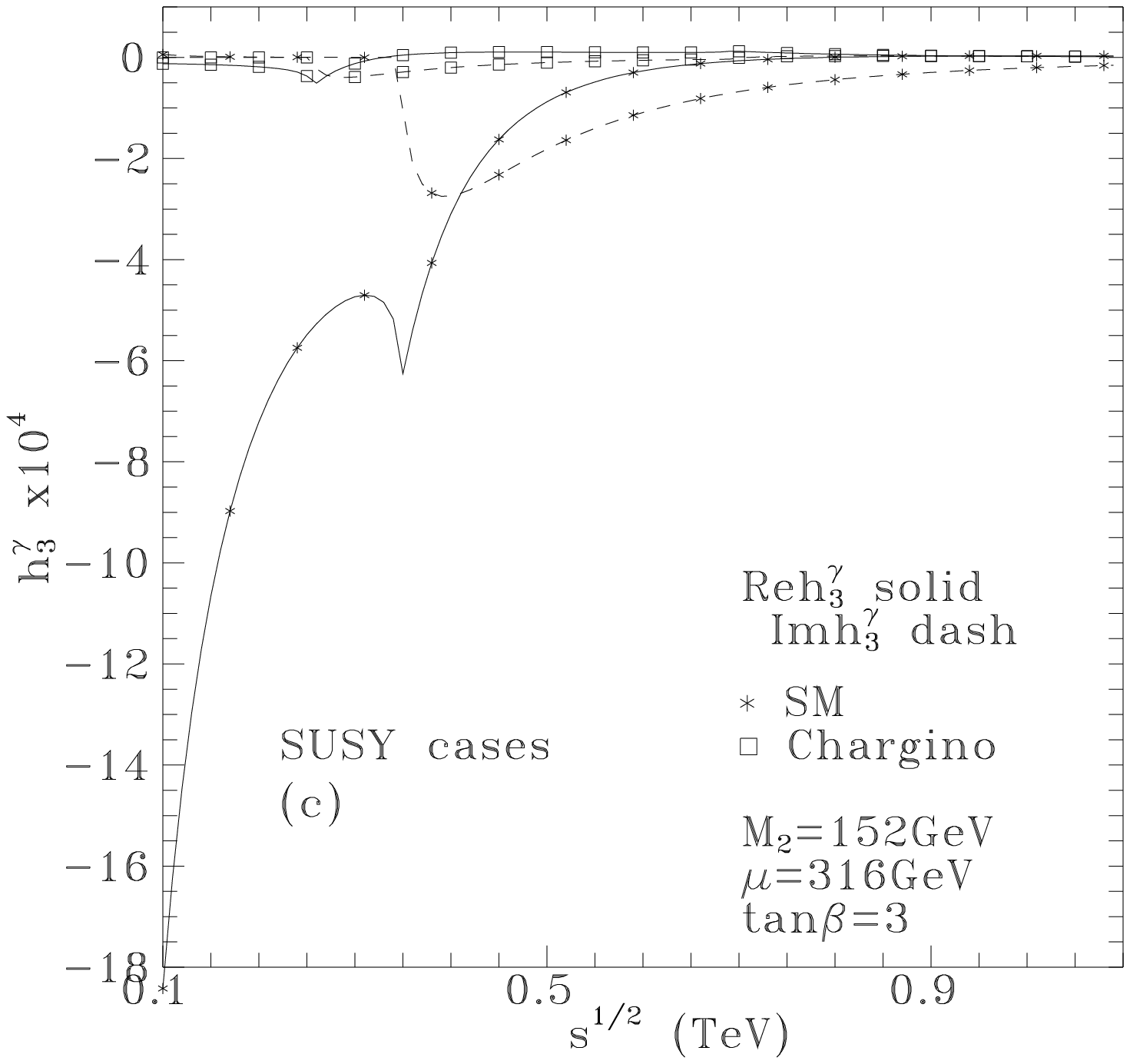,height=7.5cm}\hspace{0.5cm}
\epsfig{file=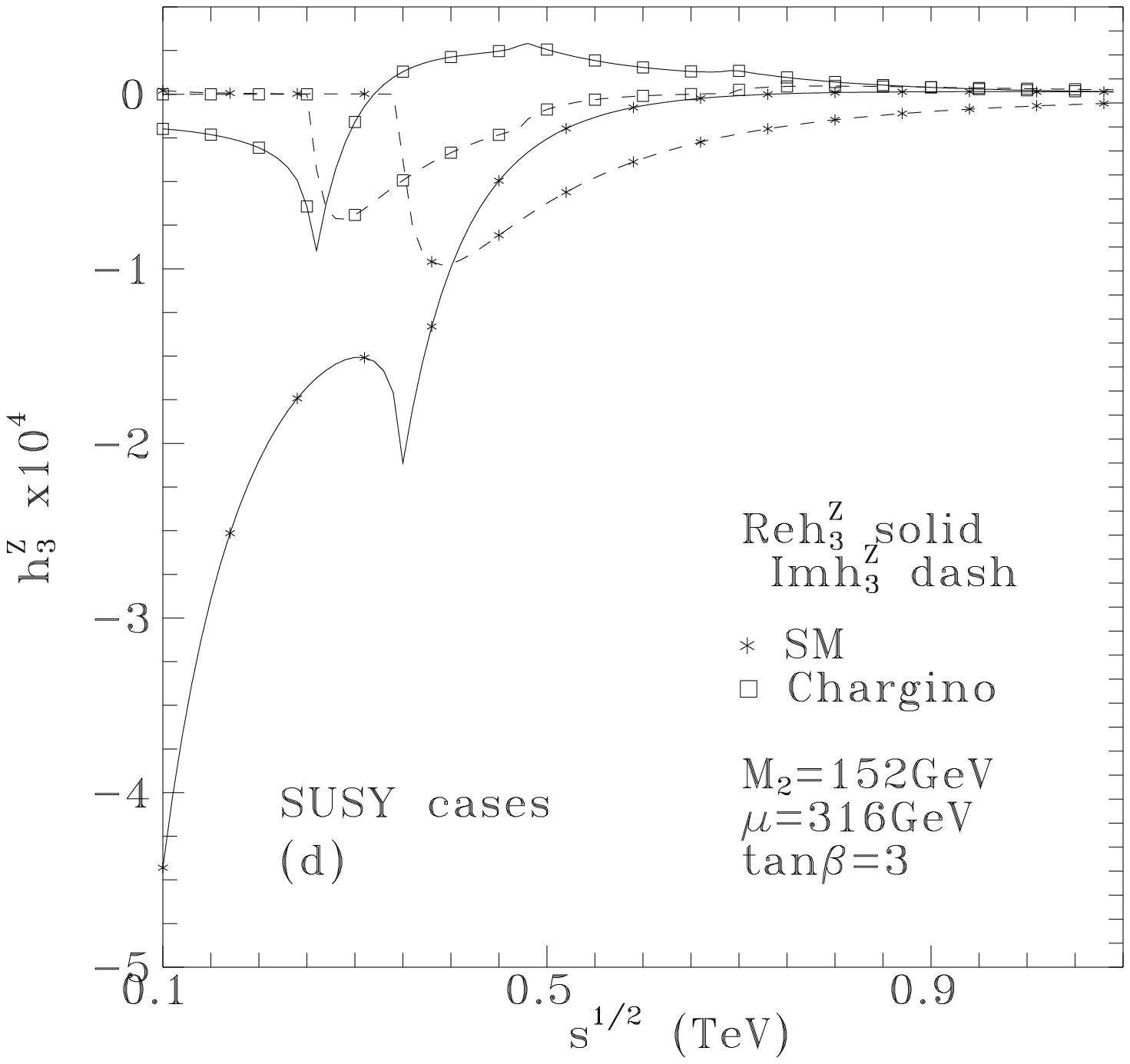,height=7.5cm}
\]
\vspace*{0.5cm}
\caption[1]{Chargino contribution for the {\bf Set 5} SUSY scenario.
For comparison the SM contribution is also shown.
The relevant parameters are indicated in the figure.}
\label{Set5-fig}
\end{figure}

\begin{figure}[p]
\vspace*{-4cm}
\[
\epsfig{file=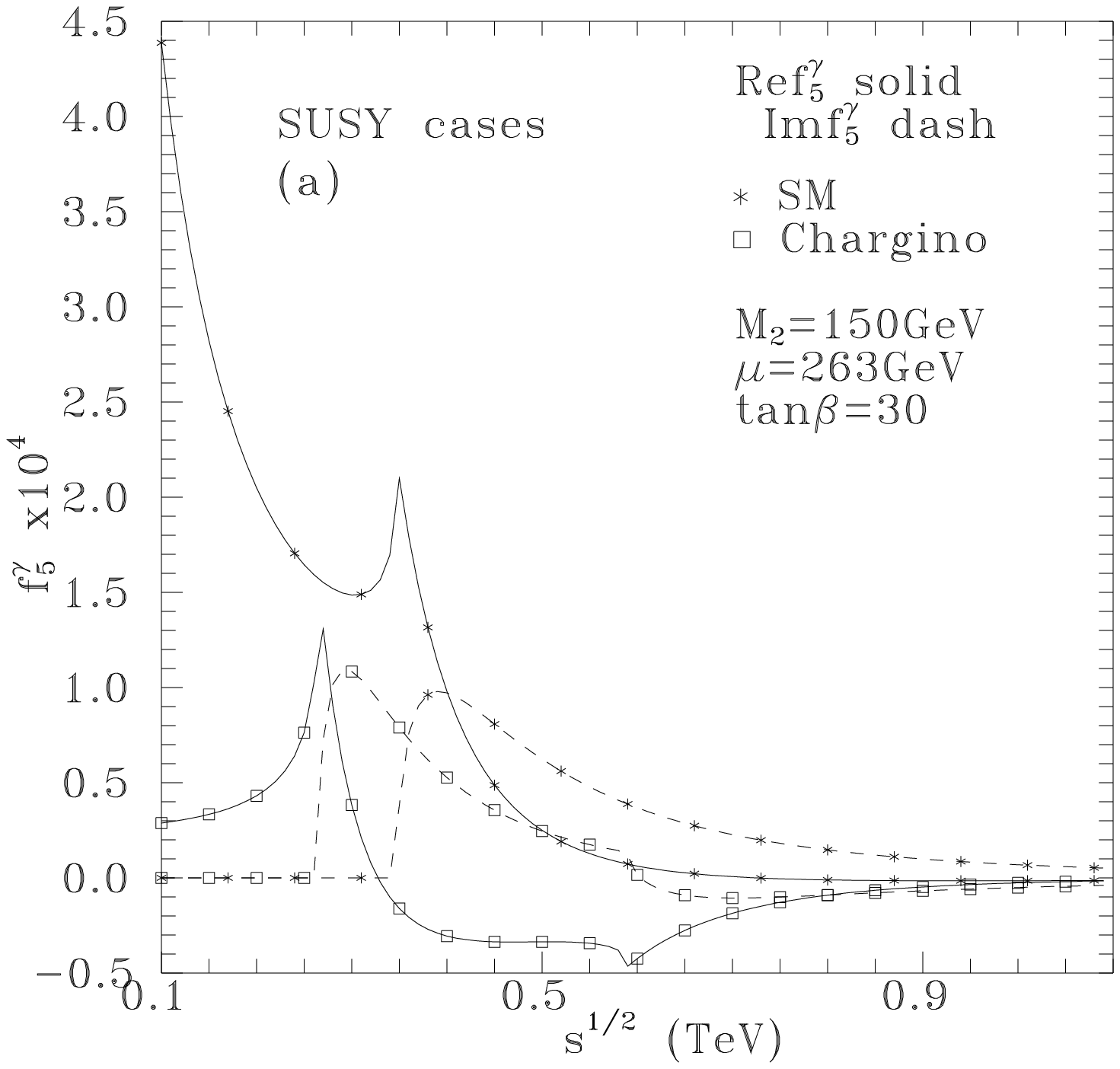,height=7.5cm}\hspace{0.5cm}
\epsfig{file=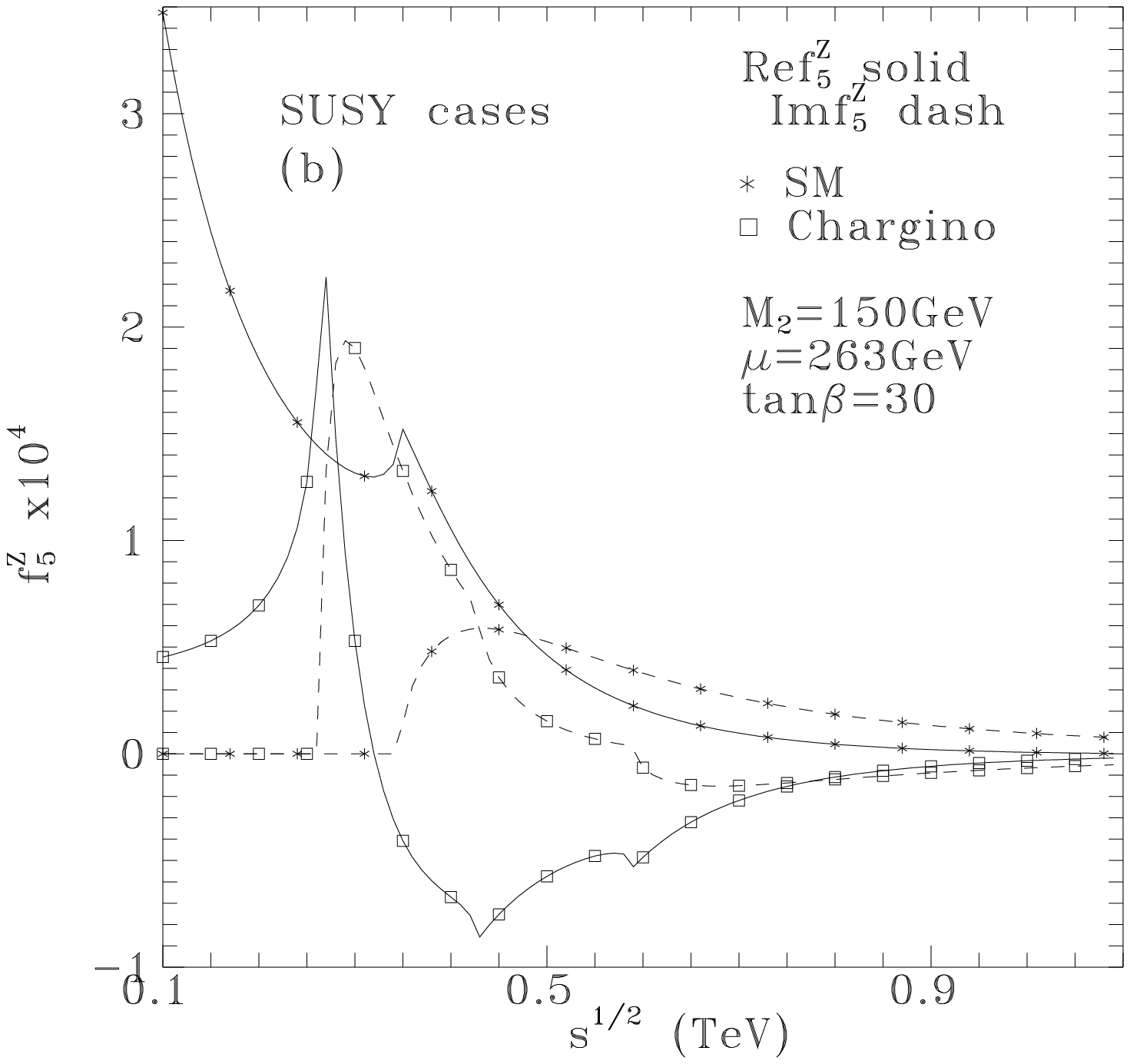,height=7.5cm}
\]
\vspace*{0.5cm}
\[
\epsfig{file=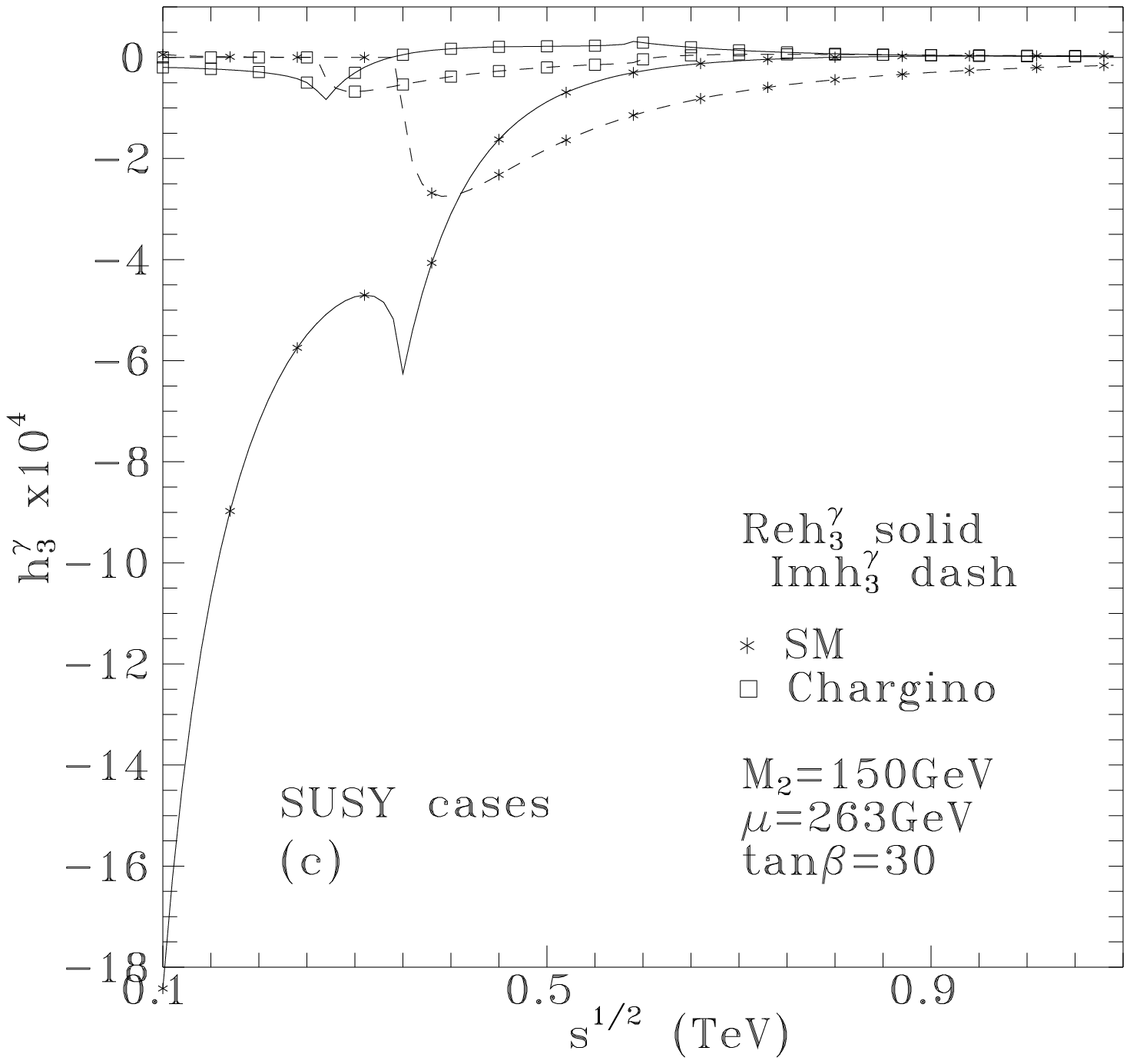,height=7.5cm}\hspace{0.5cm}
\epsfig{file=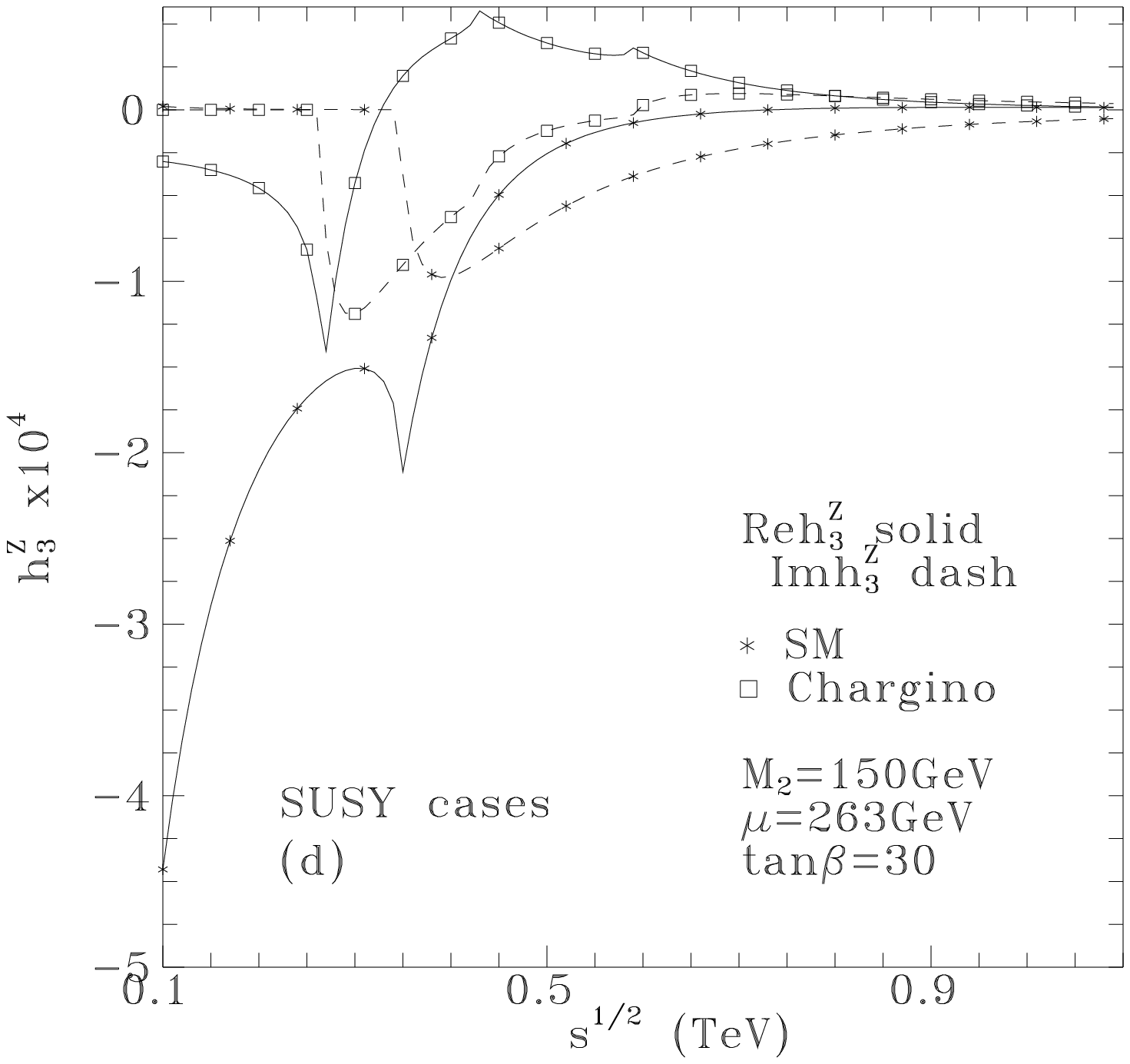,height=7.5cm}
\]
\vspace*{0.5cm}
\caption[1]{Chargino contribution for the {\bf Set 6} SUSY scenario.
For comparison the SM contribution is also shown.
The relevant parameters are indicated in the figure.}
\label{Set6-fig}
\end{figure}

Subsequently, we have computed the supersymmetric
(chargino and neutralino)
contributions in  MSSM, varying the input masses inside the
currently reasonable  ranges appearing
in Table 2 and eq.(\ref{neutralino-masses}).
The results strongly depend on the mass of the lightest SUSY state
(chargino or neutralino).
If this is
close to $100~GeV$, then the SUSY contribution
 can be comparable to or even larger (if the chargino and
neutralino effects cumulate) than the contribution from the
 top quark, or even the total  SM one; compare
 Figs.\ref{Set5-fig},\ref{Set6-fig} with Fig.\ref{SM-fig},
and the results in Table 3.
As in the SM top case, a non negligible
imaginary part is  again  generated just above the
chargino or neutralino thresholds.\par

To summarize, the "low mass" contributions, both
in SM and in MSSM, predict complex and strongly energy dependent
values for these couplings, with remarkable cusp and peak effects
which could however only be observed at a high luminosity LC. Note
also that for such small values of the couplings no effect from
the imaginary parts should be observable in  $ZZ$ or
$Z\gamma$ production cross sections, because there is no
imaginary tree level contribution
with which they could interfer, see \cite{neut}.\par

\begin{figure}[p]
\vspace*{-4cm}
\[
\epsfig{file=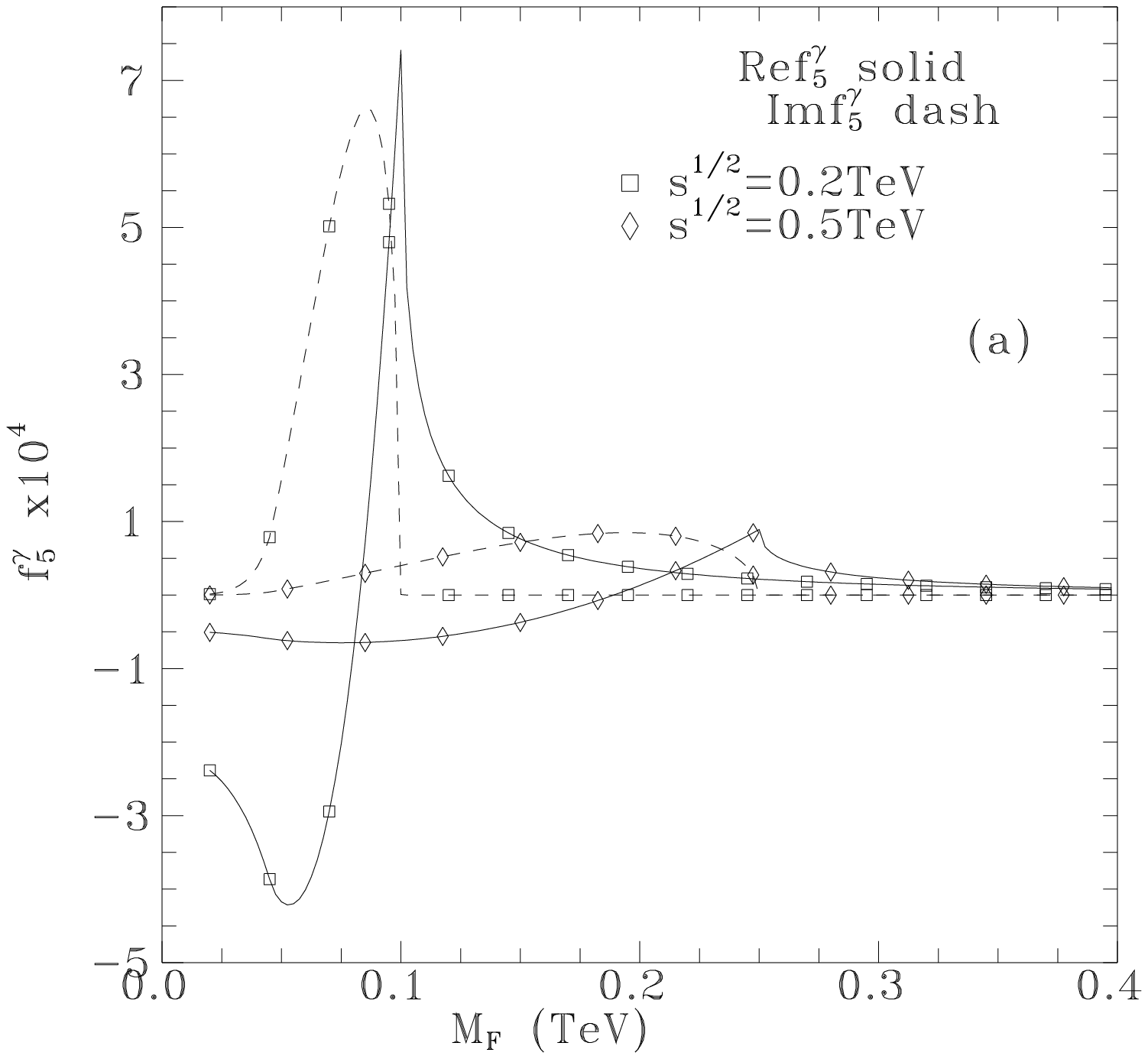,height=7.5cm}\hspace{0.5cm}
\epsfig{file=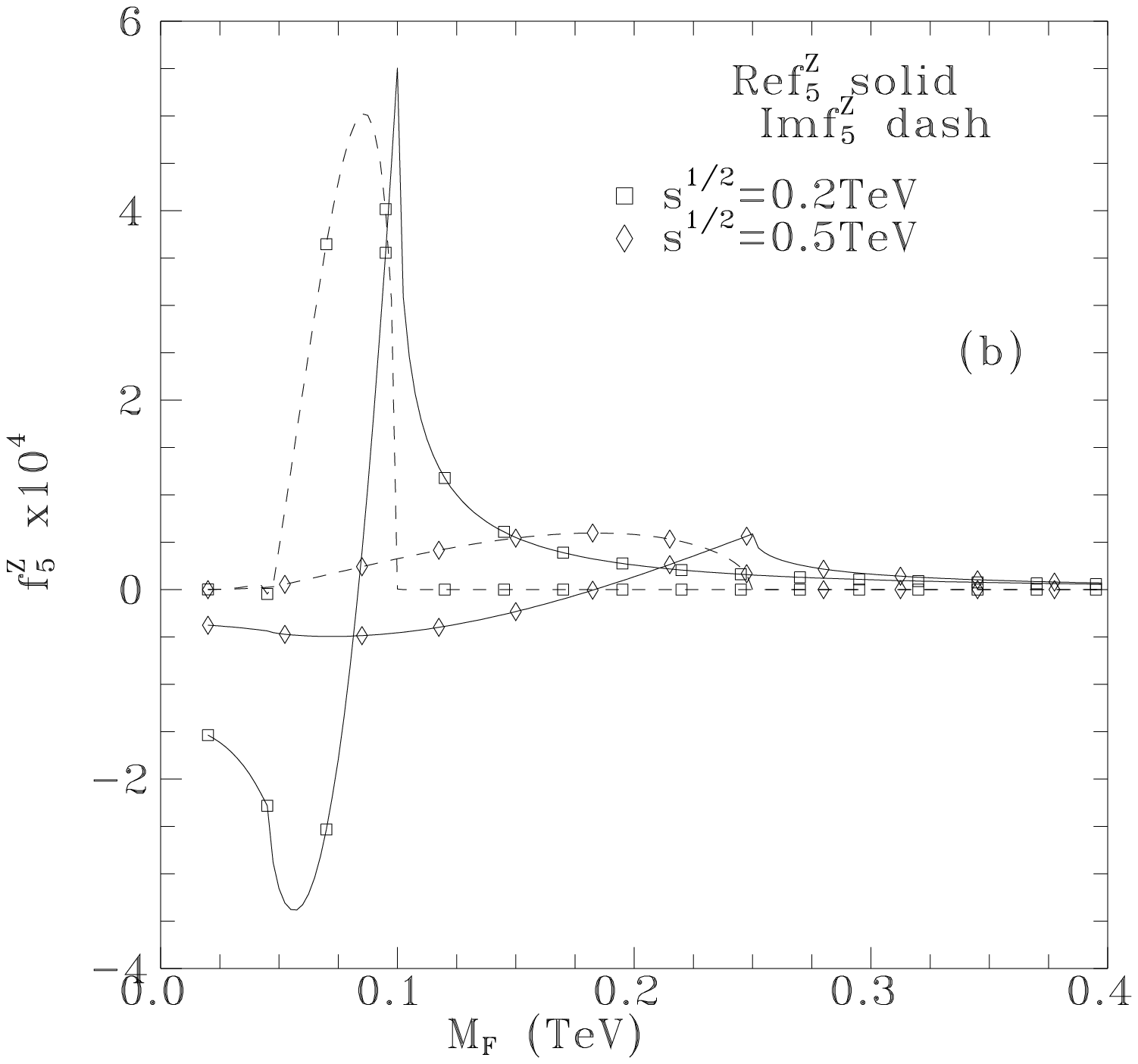,height=7.5cm}
\]
\vspace*{0.5cm}
\[
\epsfig{file=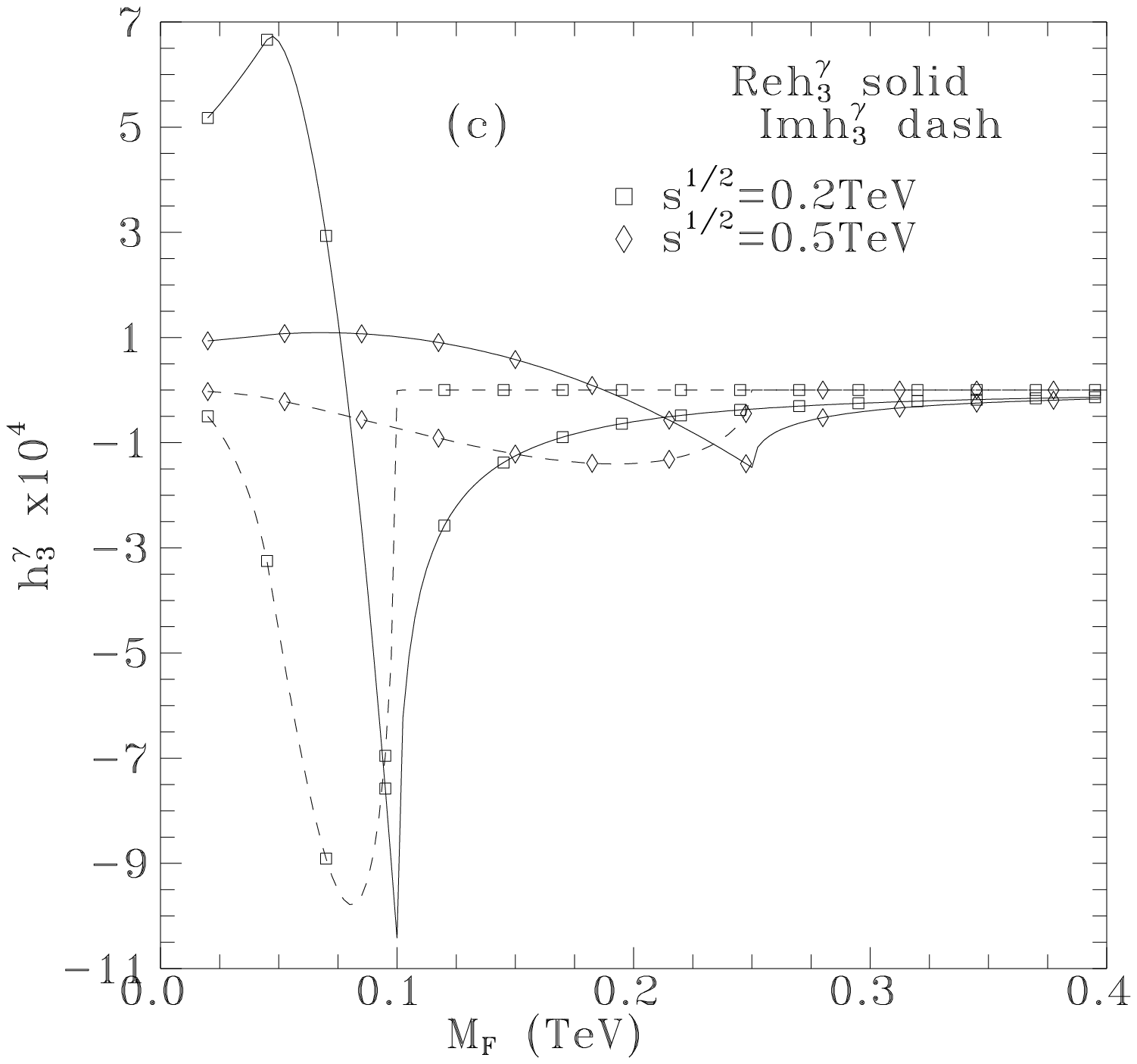,height=7.5cm}\hspace{0.5cm}
\epsfig{file=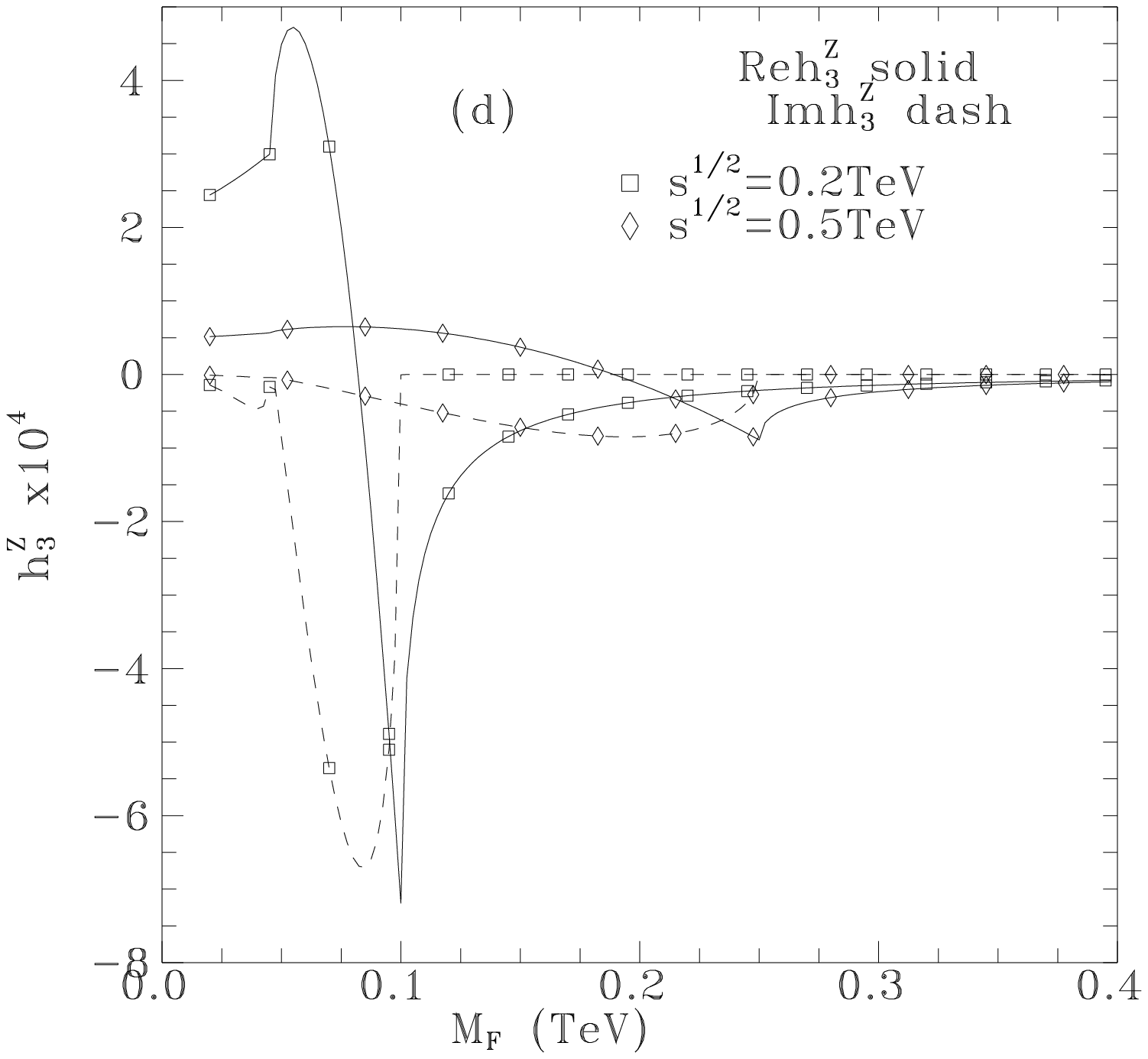,height=7.5cm}
\]
\vspace*{0.5cm}
\caption[1]{Single heavy fermion contribution to versus
$ M_F$ at $\sqrt{s_{e^-e^+}}=200$  and $500GeV$.}
\label{NPfermion-fig}
\end{figure}

To acquire a somewhat more model-independent feeling, we
have also considered the contributions of a single heavy fermion,
supposed to be the lightest one of a new physics spectrum.
Keeping its gauge couplings to photon and $Z$ as standard,
we have studied its contribution versus its mass $M_F$. Results are
presented in Fig.\ref{NPfermion-fig} for  medium values of $M_F$,
while empirical formulae have been established for higher values.
Obviously, as soon as the fermion is too heavy to be produced in pairs,
no imaginary part is generated by such NP. The real parts
decrease like $1/M^2_F$ and become of the order of $10^{-6}$ for a mass
$M_F$ in the TeV range; compare (\ref{NP-h3g}-\ref{NP-f5Z}).
So one would need either a
large colour-hypercolour factor or enhanced couplings of the heavy
fermion to the photon or to the $Z$,
in order to get observable NP contributions
of this type.\par

Finally we have discussed how high order
effects could feed the couplings
$h_4^V$, which are still vanishing at one loop.
However such contributions are always accompanied by an
additional $\alpha/4\pi$ factor, which make them totally unobservable.
Thus  the only chance to generate $h_4^V$ is  through some kind of
non perturbative contributions.
An example of such effect based on analogy with low energy
hadronic Physics and the Vector Dominance Model has been
considered. But even such a rather extreme model only leads to
\bq
\frac{h_4^V}{h_3^V}  \sim  \frac{m^2_Z}{\Lambda^2_{NP}}
~~ , \nonumber
\eq
 which for reasonable values of $\LamNP$
should render $h_4^V$  unobservable.

Our overall  conclusion is that it is very unlikely that the neutral
gauge couplings would depart \underline{in an
observable way} from SM prediction; provided of course that
 the new states couple to the
photon and  $Z$ through standard gauge couplings.
Exceptions to this statement could come, either from
low-lying (order 100 GeV) new states (for example light charginos or
neutralinos); or from enhanced photon-new fermion or $Z$-new fermion
couplings, (for example due to some resonant states of the VDM type).

\newpage

\newpage

\renewcommand{\theequation}{A.\arabic{equation}}
\renewcommand{\thesection}{A.\arabic{section}}
\setcounter{equation}{0}
\setcounter{section}{0}

{\large \bf Appendix A: Fermion loop contributions in terms of
Passarino-Veltman functions.}\\

Using the short-hand notation of the Passarino-Veltman functions
 \cite{PVHag}
\bqa
B_0(s;ij) &=& B_0(s; M_i, M_j) \\
B_Z(s; ij)& \equiv  & B_0(s;M_i, M_j)-
B_0(m^2_Z +i\epsilon;M_i, M_j) \ , \label{BZ} \\
C_{ZZ}(s;ijk) &\equiv &  C_0(\mzd, \mzd, s; M_i,M_j,M_k) \ ,
\label{CZZ} \\
C_{Z\gamma}(s;ijk) & \equiv & C_0(\mzd,0,s;M_i,M_j,M_k)
\ , \label{CZgamma}
\eqa
where $s\equiv P^2$ , and observing the symmetry relations
\bqa
B(s; ij) &=& B(s;ji)  ~ , \label{Bij-sym}\\
C_{ZZ}(s; ijk)&=& C_{ZZ}(s;kji) ~ , \label{CZZijk-sym}
\eqa
we get for the contribution of $F_j$-fermion loop
\bqa
f^{\gamma}_5(j) &=&-~\frac{N_F Q_j e^2\mzd g_{aj}g_{vj}}
{8\pi^2 s^2_Wc^2_W(s-4\mzd)^2 s}
\Bigg \{2B_Z(s;jj) \mzd (s+2\mzd)+(s-2\mzd)(s-4\mzd)
\nonumber \\
&+ & 2 C_{ZZ}(s;jjj)[M_j^2 (s-4\mzd)(s-2\mzd)+2 \mz^4 (s-\mzd)]
\Bigg \} ~~, \label{f5g-j} \\
f^{Z}_5(j)& =& -~\frac{N_F e^2\mzd g_{aj}}{16\pi^2 s^3_Wc^3_W(s-4\mzd)^2}
\Bigg \{\frac{(3 g_{vj}^2+ g_{aj}^2)(s-4 \mzd)}{3} \nonumber \\
 &+ & \frac{B_Z(s;jj)}{s-\mzd}\left [ 4 g_{aj}^2 M_j^2 (s-4\mzd)+
 (3 g_{vj}^2 +  g_{aj}^2)\mzd (s+ 2\mzd)\right ] \nonumber \\
&+& 2 C_{ZZ}(s;jjj) \left [
(g_{vj}^2+g_{aj}^2)M_j^2 (s-4\mzd)+\mz^4 (3 g_{vj}^2+g_{aj}^2)\right ]
\Bigg \} ~~, \label{f5Z-j} \\
h^\gamma_3(j) & =& \frac{N_F e^2 Q_j^2 g_{aj} \mzd }
{4\pi^2 \cw \sw (s-\mzd)}\left [ 1 +2 M_j^2 C_{Z\gamma}(s;jjj) +
\frac{\mzd}{s-\mzd} B_Z(s;jj)\right ] ~~~ , \label{h3g-j}\\
h^Z_3(j) & =& \frac{N_F e^2 Q_j g_{vj} g_{aj} \mzd }
{4\pi^2 \swd \cwd (s-\mzd)^2}\Big \{ \frac{s\mzd}{s-\mzd}~ B_Z(s;jj)
\nonumber \\
&+ & \frac{s+\mzd}{2}\, [ 2 C_{Z\gamma}(s;jjj)
M_j^2 +1 ] \Big \} ~~ , \label{h3Z-j}
\eqa
where $M_j$ is the $F_j$ mass, and $N_F$ denotes any  (colour
hypercolour) counting factor. \par

\begin{figure}[h]
\vspace*{-2cm}
\[
\epsfig{file=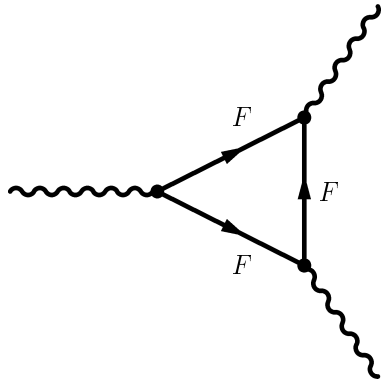,height=5cm,width=5cm}
\]
\vspace*{0cm}
\caption[2]{The fermionic triangle.}
\label{trif-fig}
\end{figure}

The mixed term contribution  arises when two different fermions,
having the same charge but mixed $ZF_1F_2$-couplings,
are running along the loop in   Fig\ref{trif-fig}.
We consider mixed $ZF_1F_2$ couplings
of the type appearing in the second line of (\ref{VheavyF}),
which arise \eg~ in the   chargino  and neutralino
cases\footnote{For simplicity we disregard the case, that three
different fermions run along the loop in Fig.\ref{trif-fig},
which could in principle only arise in the case of three light
neutralino states.}. Thus,   for the $ZZ\gamma^*$ case we obtain
\bq
f^{\gamma}_5(12) = ~ - ~\frac{N_F Q_1 e^2\mzd g_{a12}g_{v12}}
{8\pi^2 s^2_Wc^2_W(s-4\mzd)^2 s} R_{ZZ\gamma^*} ~~ ,
\label{f5g-12}
\eq
where
\bqa
&& R_{ZZ\gamma^*}=
4(s-\mzd)(M_2^2-M_1^2)[B_0(s;11)-B_0(s;22)]+2(s-2\mzd)(s-4\mzd)
\nonumber \\
&&+ 2\mzd (s+2\mzd)[B_0(s;11)+B_0(s;22)-2 B_0(\mzd+i\epsilon;12)]
\nonumber \\
&& +4 (s-\mzd)(M_1^4+M_2^4+\mz^4 -2M_1^2M_2^2-2M_1^2\mzd
-2M_2^2\mzd) [C_{ZZ}(s;121)+C_{ZZ}(s;212)]
\nonumber \\
&& + 2 s (s+2\mzd)[M_2^2 C_{ZZ}(s;121)+M_1^2 C_{ZZ}(s;212)]
~~ , \label{RZZg}
\eqa
and $Q_1=Q_2$ is the common charge of $F_1,~ F_2$.  \par

Correspondingly for the   $ZZ Z^*$ case we get
\bqa
&& f^{Z}_5(12) = -~\frac{N_F e^2\mzd }{16\pi^2 s^3_Wc^3_W(s-4\mzd)}
\Bigg \{[g_{a1}(g_{a12}^2+g_{v12}^2)+2 g_{a12}g_{v12}g_{v1}]R_a
\nonumber \\
&&+  [g_{a1} (g_{a12}^2+g_{v12}^2)-2 g_{a12} g_{v12} g_{v1}] R_b +
g_{a1}(g_{a12}^2- g_{v12}^2) R_c ~+~ (1 \leftrightarrow 2) \Bigg
\} ~ , \label{f5Z-12}
\eqa
where
\bqa
&&R_a  = 1 +~ \frac{2(M_1^2-M_2^2)(s+2\mzd)}{s (s-4\mzd)} [B_0(\mzd;
12)-B_0(\mzd ;11)]~ \nonumber \\
&& -~\frac{1}{(s-\mzd)(s-4\mzd)} \Bigg \{
\frac{C_{ZZ}(s ;112)}{s} \Big [-2 M_1^2 s^3 -s^2 \Big (
M_1^2 (M_1^2-M_2^2) -
\nonumber \\
&& \mzd (7M_1^2+3M_2^2-4\mzd) \Big )
 + 4 s \mzd (M_1^2 M_2^2 -2 M_1^2\mzd -M_2^4+\mz^4) +4\mz^4
(M_1^2-M_2^2)^2 \Big ] \nonumber \\
&&  - C_{ZZ}(s; 121) \Big [
s^2 M_2^2+s (M_1^4-3 M_1^2 M_2^2- 3M_1^2 \mzd +2M_2^4-2 M_2^2\mzd
+2\mz^4) \nonumber \\
&& + 2 \mzd (M_1^4-M_2^4 +2M_2^2 \mzd -\mz^4) \Big ]
+B_Z(s;11) [(M_1^2-\mzd)(s+2\mzd)\nonumber \\
&& -2M_2^2 (s-\mzd) ]
 +B_Z(s;12) [-3M_1^2 (s-2 \mzd) +(M_2^2 -2\mzd)(s +2\mzd) ]
\Bigg \} ~,  \label{Ra} \\
&& R_b = \frac{M_1^2}{s-\mzd} \Bigg \{
(M_1^2-M_2^2
-\mzd)[C_{ZZ}(s;112)-C_{ZZ}(s;121)] \nonumber \\
&& +(s-\mzd)C_{ZZ}(s;112) +B_Z(s;11) +2 B_Z(s;12) \Bigg \} ~,
\label{Rb} \\
&& R_c = \frac{M_1 M_2}{s-\mzd} \Bigg \{
(2M_1^2-2M_2^2 +s )[C_{ZZ}(s;112)-C_{ZZ}(s;121)] \nonumber \\
&& +2(s-\mzd)C_{ZZ}(s;121) +2B_Z(s;11) +4 B_Z(s;12) \Bigg \}
~ . \label{Rc}
\eqa

Finally for the $Z\gamma Z^*$ mixed case we have
\bqa
h^Z_3(12) & =& \frac{N_F e^2 Q_1 g_{v12} g_{a12} \mzd }
{4\pi^2 \swd \cwd (s-\mzd)^2}\Big \{ \frac{2 s\mzd}{s-\mzd}~ B_Z(s;12)
\nonumber \\
&+ & (s+\mzd)[ M_2^2 C_{Z\gamma}(s;122)+ M_1^2 C_{Z\gamma}(s;211)
 +1 ] \Big \} ~~ . \label{h3Z-12}
\eqa

We note that there is no mixed contribution for $h_3^\gamma$.

\newpage
\renewcommand{\theequation}{B.\arabic{equation}}
\renewcommand{\thesection}{B.\arabic{section}}
\setcounter{equation}{0}
\setcounter{section}{0}

{\large \bf Appendix B: Fermion loop contributions in terms of
Feynman integrals.}\\

 The contributions of a single $F_j$-fermion loop
in terms of Feynman integrals are:
\bq
f^{\gamma}_5(j)=
N_F{e^2Q_jg_{vj}g_{aj}\over4\pi^2s^2_Wc^2_W}~I^{\gamma}_5
\label{f5g} ~ ,
\eq
\bq
f^{Z}_5(j)=-~N_F{e^2g_{aj}\over96\pi^2s^3_Wc^3_W}~
([g_{aj}^2+3g^2_{vj}]~I^{Z1}_5+
[g_{vj}^2-g^2_{aj}]~I^{Z2}_5) ~ ,
\label{f5Z}
\eq
\bq
h^{\gamma}_3(j)=-~N_F{e^2Q^2_jg_{aj}\over2\pi^2s_Wc_W}~I^{\gamma}_3
~ , \label{h3g}
\eq
\bq
h^Z_3(j)=-~N_F{e^2Q_jg_{vj}g_{aj}\over4\pi^2s^2_Wc^2_W}~I^{Z}_3
 ~ ,\label{h3Z}
\eq
\noindent
with
\bq
I^{\gamma}_5=
\int^1_0dx_1\int^{1-x_1}_0dx_2 ~{x_1(1-x_1-x_2)m^2_Z
\over D_{ZZ}(s)} ~ ,
\label{I5g}
\eq
\bqa
I^{Z1}_5=&&{6m^2_Z\over s-m^2_Z} \int^1_0dx_1\int^{1-x_1}_0dx_2~
 \cdot \nonumber\\
&&\{{[(1-x_1-x_2)(m^2_Zx^2_2-sx^2_1)+x_2(1-3x_1)(sx_1+m^2_Z(2x_2-1))]
\over D_{ZZ}(s)}\nonumber\\
&&-~{m^2_Z[(1-x_1-x_2)(x^2_2-x^2_1)+x_2(1-3x_1)(x_1+2x_2-1)]
\over D_{ZZ}(m^2_Z)}\} ~ , \label{IZ1}
\eqa
\bq
I^{Z2}_5={6M^2_Fm^2_Z\over s-m^2_Z}\int^1_0dx_1\int^{1-x_1}_0dx_2~
{(1-3x_1)
\over D_{ZZ}(s)} ~ , \label{IZ2}
\eq
\noindent
where
\bq
D_{ZZ}(s)\equiv M^2_j+sx_1(x_1+x_2-1)+m^2_Zx_2(x_2-1) ~ ,
\eq
and
\bq
I^{\gamma}_3=
\int^1_0dx_1\int^{1-x_1}_0dx_2 ~{x_1(1-x_1-x_2)m^2_Z
\over D_{Z\gamma}(s)} ~ , \label{I3g}
\eq
\bq
I^{Z}_3={m^2_Z\over s-m^2_Z}\int^1_0dx_1\int^{1-x_1}_0dx_2~
{(sx_1-m^2_Zx_2)(1-x_1-x_2)\over D_{Z\gamma}(s)} ~ ,
\label{I3Z}
\eq
\noindent
where
\bq
D_{Z\gamma}(s)\equiv M^2_j+sx_1(x_1+x_2-1)+m^2_Zx_2(x_1+x_2-1)
~ .
\eq
\vspace{0.5cm}

 The mixed contributions due to two different fermions (with the same
charge) around the loop are:
\bq
h^Z_3(12)=-~N_F{e^2Q_1g_{v12}g_{a12}\over4\pi^2s^2_Wc^2_W}~I'^{Z}_3
~+~ (1 \leftrightarrow 2) ~ , \label{h3Zm}
\eq
where  $I'^{Z}_3$ is given by (\ref{I3Z}) with $D_{Z\gamma}(s)$
replaced by
\bq
D'_{Z\gamma}(s)\equiv M^2_1+(M^2_2-M^2_1)(1-x_1-x_2)+
sx_1(x_1+x_2-1)+m^2_Zx_2(x_1+x_2-1) ~ .
\label{Dzgp}
\eq

Correspondingly
\bq
f^{\gamma}_5(12)=
N_F{e^2Q_1g_{v12}g_{a12}\over4\pi^2s^2_Wc^2_W}~I'^{\gamma}_5
~+~ (1 \leftrightarrow 2)  ~ ,\label{f5gm}
\eq
where
$I'^{\gamma}_5$ is given by the same expression as in (\ref{I5g})
with $D_{ZZ}(s)$ replaced by
\bq
D'_{ZZ}(s)\equiv M^2_1+(M^2_2-M^2_1)x_2+
sx_1(x_1+x_2-1)+m^2_Zx_2(x_2-1) ~ . \label{Dzzp}
\eq

Finally,
\bqa
&&f^{Z}_5(12)=-~N_F{e^2\over96\pi^2s^3_Wc^3_W}~
([g_{a1}(g_{a12}^2+g^2_{v12})+2g_{v1}g_{a12}g_{a12}]~I^{Za}_5
+M^2_1[g_{a1}(g_{a12}^2+g^2_{v12})\nonumber\\
&&-2g_{v1}g_{a12}g_{v12}]~I^{Zb}_5
+2M_1M_2g_{a1}(g_{v12}^2-g^2_{a12})~I^{Zc}_5)
~+~ (1 \leftrightarrow 2)~ ,\label{f5Zm}
\eqa
\noindent
with
\bqa
&&I^{Za}_5={3m^2_Z\over s-m^2_Z}\int^1_0dx_1\int^{1-x_1}_0dx_2 ~
\{[sx_1(1-x_1-x_2)(x_2-1)+m^2_Zx^2_2(1-x_2)].
 \nonumber \\
&&.({1\over D_1}+{2\over D_2})+(1-3x_2)(lnD_1+2lnD_2)\}
\eqa
\bq
I^{Zb}_5={3m^2_Z\over s-m^2_Z}\int^1_0dx_1\int^{1-x_1}_0dx_2
\{{1-x_1\over D_1}-{2x_2\over D_2}\}
\eq
\bq
I^{Zc}_5=
{3m^2_Z\over s-m^2_Z}\int^1_0dx_1\int^{1-x_1}_0dx_2
\{{2x_2\over D_1}-{2(1-2x_2)\over D_2}\}
\eq

and
\bqa
&&D_1=M^2_1+(M^2_2-M^2_1)x_2+sx_1(x_1+x_2-1)+m^2_Zx_2(x_2-1)
~ , \nonumber\\
&&D_2=M^2_1+(M^2_2-M^2_1)x_1+sx_1(x_1+x_2-1)+m^2_Zx_2(x_2-1)
~ , \label{Dp}
\eqa

Some of the unmixed parts of these expressions can be derived from the
results obtained in \cite{BBCD}. We have checked by doing a
direct numerical integration, that they agree with the
numerical results obtained from the Passarino-Veltman
expressions and the FF-package \cite{FFpack}.\par

It is easy, from the above analytic expressions, to derive
the asymptotic expressions given in eq.(\ref{h3Zs}-\ref{h3gs})
and (\ref{f5Zmodel}-\ref{h3gmodel}).\par

For comparison we also give the fermion doublet  contribution to the
$\gamma WW$ and $ZWW$ anapole couplings defined in
(\ref{ZWW-anapole}).  Denoting the "up" and "down" fermions
as $F$ and $F'$ and defining their Z-couplings as in
(\ref{VheavyF}), we get
\bqa
z_Z& = &-~N_F{e^2\over32\pi^2s_Wc_W}([(g_{vF}+g_{aF}]I_F
+[(g_{vF'}+g_{aF'}]I_{F'}) ~ ,\label{zZFFp} \\
z_{\gamma}& = & -~N_F{e^2\over16\pi^2s^2_W}(Q_F I_F+Q_{F'}I_{F'})
~ , \label{zgFFp}
\eqa
\noindent
where
\bqa
 &&I_F= \int^1_0dx_1\int^{1-x_1}_0dx_2 \cdot \nonumber \\
&&
~ {x_1(1-x_1-x_2)m^2_W
\over M^2_F+sx_1(x_1+x_2-1)+m^2_Wx_2(x_1+x_2-1)+(M^2_{F'}-M^2_F)x_2}
\label{IF}
\eqa
\noindent
and $I_{F'}=I_F(M^2_F\to M^2_{F'})$.
In the case of a degenerate doublet $M_F=M_{F'}$, with standard
 couplings, this result
simplifies to
\bq
z_Z=-~{s_W\over c_W} z_{\gamma}=N_F{e^2\over48\pi^2s_Wc_W}I_W
~ , \label{zZFDq}
\eq
\noindent
for a doublet of quarks, and
\bq
z_Z=-~{s_W\over c_W} z_{\gamma}=-N_F{e^2\over16\pi^2s_Wc_W}I_W
~ , \label{zZFDl}
\eq
\noindent
for a doublet of leptons,
with
\bq
I_W=\int^1_0dx_1\int^{1-x_1}_0dx_2~ {x_1(1-x_1-x_2)m^2_W
\over M^2_F+sx_1(x_1+x_2-1)+m^2_Wx_2(x_1+x_2-1)} ~ ,
\label{IFD}
\eq
\noindent
which, for $M_F \gg m_W$, leads to
eq.(\ref{zVgmodelq},\ref{zVgmodell}).\par
Note that for a completely degenerate standard family
(a doublet of lepton and a doublet of coloured quarks)
the total contribution vanishes.

\clearpage
\newpage

\clearpage

\newpage
\clearpage

\newpage
\clearpage

\newpage
\clearpage

\newpage
\clearpage

\end{document}